# predictSLUMS: A new model for identifying and predicting informal settlements and slums in cities from street intersections using machine learning

Mohamed R. Ibrahim[1], Helena Titheridge[2], Tao Cheng[3] and James Haworth[4]

Department of Civil, Environmental and Geomatic Engineering, University College London (UCL)

[1]mohamed.ibrahim.17@ucl.ac.uk, [2]h.titheridge@ucl.ac.uk, [3]tao.cheng@ucl.ac.uk, [4]j.haworth@ucl.ac.uk

*Abstract*— Identifying current and future informal regions within cities remains a crucial issue for policymakers and governments in developing countries. The delineation process of identifying such regions in cities requires a lot of resources. While there are various studies that identify informal settlements based on satellite image classification, relying on both supervised or unsupervised machine learning approaches, these models either require multiple input data to function or need further development with regards to precision. In this paper, we introduce a novel method for identifying and predicting informal settlements using only street intersections data, regardless of the variation of urban form, number of floors, materials used for construction or street width. With such minimal input data, we attempt to provide planners and policy-makers with a pragmatic tool that can aid in identifying informal zones in cities. The algorithm of the model is based on spatial statistics and a machine learning approach, using Multinomial Logistic Regression (MNL) and Artificial Neural Networks (ANN). The proposed model relies on defining informal settlements based on two ubiquitous characteristics that these regions tend to be filled in with smaller subdivided lots of housing relative to the formal areas within the local context, and the paucity of services and infrastructure within the boundary of these settlements that require relatively bigger lots. We applied the model in five major cities in Egypt and India that have spatial structures in which informality is present. These cities are Greater Cairo, Alexandria, Hurghada and Minya in Egypt, and Mumbai in India. The predictSLUMS model shows high validity and accuracy for identifying and predicting informality within the same city the model was trained on or in different ones of a similar context.

*Keywords*— Machine Learning; Slums; Informal Settlements; Complexity; Spatial network; Spatial statistics; Neural Networks; Egyptian cities

## 1. INTRODUCTION

### 1.1 Overview

It is evident that most of the world population will either be born in cities or will move to cities In 2015, the total urban population of the world exceeded the rural population for the first time and this trend is set to continue (UN-Habitat, 2007, 2016). As a result, many cities face challenges with accommodating this rapid urban growth.

Globally, one in three city residents is a slum dweller. This presents a major challenge for urban housing as most of these settlements are not yet seen in official governmental maps (Montgomery, 2008; Roy, Lees, Palavalli, Pfeffer, & Sloot, 2014; UN-Habitat, 2007). This adds challenges, first, for understanding the dynamics of cities in the global south. Second, it remains ambiguous how people select a particular location to create their homes and later their settlements (Ibrahim, 2017). Last, this complicates the process of diagnosing and monitoring cities to cope with the necessity of providing adequate services and enhancing the living conditions of urban dwellers.

Within cities development in the global south, informality has taken various shapes and forms. There are different ways of defining them: *"Ashwa'iat*[1]*"* of Egypt, *"Favelas*[2]*"* of Brazil, *"Campamentos*[3]*"* of Chile, or others elsewhere. Although these pinpoint similar informal regions within cities*,* they retain subtle social and spatial features that make each term a unique identifier for its local context, which are not necessarily perceived as slums but rather a diversity of *'unofficially planned areas'*. With such a wide spectrum of informal regions across the globe, the current knowledge gap remains in addressing such a diversity of informal regions with 'a unified global method' (Mahabir, Croitoru, Crooks, Agouris, & Stefanidis, 2018), that can cope with the process of rapid urbanisation and enables it to be used by policy-makers and planners in developing countries, where availability of data can be a major issue.

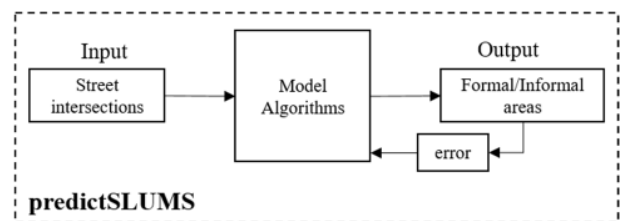

FIG. 1
MODEL CONCEPT

### 1.2 predictSLUMS concept and features

In this article, we introduce an unprecedented model for identifying and predicting informal settlements that relies only on street intersections data. The currently proposed predictSLUMS is a 'predictive-static model'. Unlike satellite image classification of slums that are more likely to be context based study, the models aims to identify the complex urban system of informal regions and slums in different cities. Its algorithms deal with informal settlements from a generic view that identifies informal regions in various contexts globally, regardless of the precise urban form, number of floors, street width, or the material used for construction. It identifies and predicts informal regions in accordance with data on delineated informal areas. It also predicts informality in regions that are not yet labelled as such due to the absence of data, by relying on understanding the configuration of informal settlements in other contexts where labelled data can be used for training the model.

---

[1] Literally, random or unplanned areas, including slums (Ibrahim, 2017; O'Donnell, 2010).

[2] Literally, chanty town or a slum (Novaes, 2014).

[3] Literally, camp or tent cities. Also known as mushroom towns (Salcedo, 2010).

Fig. 1 illustrates the basic idea of the predictSLUMS model. The model inputs street intersections data and outputs the location of informal regions in a city; the error can be back-propagated. The model output resolution is a lattice of a grid size (100m x 100m), where each cell of the grid is identified as either formal or informal. The key features of the model are:
 a. The model relies on one source of input data to function, enabling it to be used by planners and policy-makers in the global south;
 b. One model can be fitted to different cities;
 c. By training the model in one city, the model can predict informal and slum areas in a different city of a similar urban context and structure.

*1.3 Paper structure*

The paper is structured as follows: Section 2 gives a brief review of urban modelling and the detection of informal settlements. Section 3 presents the methodology and algorithms of the predictSLUMS model. Section 4 describes some cases where the model has been applied and the data used. Section 5 gives the main findings of the predictSLUMS simulation. Section 6 discusses these results, their relevancy, and the limitations of the model. Last, a summary and some conclusions are given in Section 7.

## 2. THE STATE OF URBAN MODELLING AND SLUMS DETECTION: A REVIEW

Significant advances in urban modelling have been made in recent years. For example, by perceiving cities as complex systems, it has been shown that the forms and shapes of cities follow similar developmental paths based on scaling laws and fractal dimensions that allow a bottom-up evolving of geometry within a temporal scale (Batty, 2008; Bettencourt, 2013; Cottineau, Hatna, Arcaute, & Batty, 2017; Isalgue, Coch, & Serra, 2007; Kühnert, Helbing, & West, 2006). Batty, Xie, & Sun (1999) have introduced a theoretical dynamic model that simulates the various types of urban sprawl in a city. Patel, Crooks, & Koizumi (2012, 2018) introduced an agent-based model that explores the formation of slums in Ahmedabad in India. However, within the various attempts of spatial modelling of urban expansion and residential location choice, identifying informal zones within cities remains a missing attribute in many urban models (Berberoğlu, Akın, & Clarke, 2016; Feldman et al., 2007; O'Donoghue, Morrissey, & Lennon, 2014; Shafizadeh-Moghadam, Asghari, Tayyebi, & Taleai, 2017; Tian et al., 2016; Waddell et al., 2003).

What makes the detection of informality a complex process is the lack of general definition of what informal settlements are. The subtlety of differences in spatial features between formal and informal settlements, in many cases, complicates the classification process among these merged regions (Hofmann, Taubenböck, & Werthmann, 2015).

There is a wide range of literature that introduces methods of image classification for various applications of remote sensing, relying on both supervised and unsupervised machine learning (Li, Wang, Li, & Chen, 2017; Liu et al., 2016; Lüscher, Weibel, & Burghardt, 2009; Ma et al., 2017; Patino & Duque, 2013; Sharma, Liu, Yang, & Shi, 2017; Tan, Hu, Li, & Du, 2015; W. Zhang et al., 2017; Zhu, Li, Hu, & Wu, 2017), that is also adopted among different scholars for identifying informal and slum areas (Kuffer, Pfeffer, & Sliuzas, 2016). For instance, Kohli, Sliuzas, & Stein (2016) introduced a method for slums detection based on satellite image classification by applying experts' knowledge of the morphology of local slums area in Pune, India. They also implemented an object-based image detection for slum areas in Ahmedabad, Cape Town and Nairobi, relying on complementary data acquired from questionnaires (Kohli, Stein, & Sliuzas, 2016). Similarly, an image classification approach for identifying slums in Accra, Ghana (Engstrom et al., 2015), in Hyderabad, India (Kit, Lüdeke, & Reckien, 2012), and in La Paz, Kabul, Kandahar, and Caracas based on supervised machine learning (Graesser et al., 2012). Wurm, Taubenböck, Weigand, & Schmitt (2017) relied on Synthetic Aperture Radar data for mapping and identifying slum areas in Mumbai, using supervised machine learning approach for classification. Data captured from Unmanned Aerial Vehicles was used to identify informality in Maldonado, Uruguay and Kigali, Rwanda (Gevaert, Persello, Sliuzas, & Vosselman, 2017). Optical Spaceborne data were used to classify informality in arid areas (Stasolla & Gamba, 2007). Agent-based modelling has been an approach for detecting the growth of the informal settlements in Dar-es-Salaam based on vector-based data represented in building blocks (Augustijn-Beckers, Flacke, & Retsios, 2011). Methods of detecting informal settlements may vary. However, what is clear is that all these approaches discussed above require multiple and sophisticated data to which access may be limited for many less developed countries. This highlights the necessity of a unified method that can be used globally (Mahabir et al., 2018).

On the other hand, flows and networks systems are indispensable for understanding spaces in cities (Batty, 2013). Since the emergence of network theory, the spatial network has been an important means of data for urban studies (Barthélemy, 2011; Zhong, Arisona, Huang, Batty, & Schmitt, 2014). For example, Boeing (2017a) used street network data to analyse the complexity of urban morphology in the US cities. Arcaute et al., (2016) used them to identify the hierarchy of cities and regions in the U.K. relying on percolation theory. Nevertheless, population growth in cities was defined based on the same means of data (Arcaute et al., 2014; Masucci, Arcaute, Wang, et al., 2015, 2015; Masucci, Arcaute, Hatna, Stanilov, & Batty, 2015). Accordingly, the availability of the spatial network data and their significant impact on the studies of urban morphologies raise more questions of how can we rely on them to understand more about the fundaments of cities development when slums and informal settlements are present.

## 3. METHODOLOGY

Hypothetically, informal settlements are characterised by small subdivided lots that allow the incremental process of self-build housing. These regions are characterised by the paucity of large public spaces and services that require relatively larger lots than those assigned for housing. These leave the layout of such a settlement filled in with a random pattern of nearly equal small subdivided lots relative to their local formal neighbouring settlements, regardless of the urban form, number of floors, the materials used for construction, or the width of the local streets. Consequently, understanding the interrelation between street intersections of a region in comparison to neighbouring regions can provide a classification of planned and non-planned regions. Thus, it can define informal settlements in different context globally. Based on this hypothesis, this paper addresses two research questions. *First, how can we identify the status of the built-environment (formal, or informal) using street network data? Second, is it possible to predict informal areas and slums in a city by understanding housing informality in other cities of similar context?*

The model algorithms are divided into two phases; the model deals with each phase consecutively. The first phase deals with clustering indices and variables selection. It shows the association and the likelihood of the computed variables with the informal settlements that may or may not fully represents the actual settlements, whereas the second phase of the model predicts the actual formal/informal regions. Fig 2 shows the general architecture of predictSLUMS. The first phase focuses on computing variables that are more likely to be associated with informality relying on geospatial analysis. The second phase deals with the prediction aspects of the model. It relies on the architecture of Artificial Neural network (ANN) to output the informal regions for the studied area.

The figure also shows the transition of the spatial representation and resolution of the model. It shows the transition from clouds of points that represents the street intersections towards the final resolution of the model of a grid size (100m x 100m). There are two crucial aspects to be noted. First, after computing the variables from the street intersections in the first phase, the results of these variables create a grid cell with centroids, in which they no longer represent the actual street intersections points. The prediction of the ANN model is then computed using these centroids. This allows better accessibility and greater freedom for using the predicted results of the model for any further research or urban simulation that may be conducted in various programming languages.

In this study, we refer to the street intersections as *'incident points'* as they do not carry any feature attributes apart from their coordinates.

### 3.1 Phase I: Identification
#### 3.1.1 Clustering and computing variables

To identify homogenous clusters that may represent informal settlements in cities, three methods were analysed. First, we conducted Average Nearest Neighbour (Nn) analysis to test whether these incident points are significantly clustered or dispersed (Clark & Evans, 1954). This analysis computes the average distance between each point and its nearest neighbours. If the computed average distance is less than the hypothetical randomly distributed distance, then the points are significantly clustered. Contrary, if the observed distance is below the expected hypothetical distance, then the points are dispersed. Nn is calculated as:

$$Nn = \frac{\overline{D}_o}{\overline{D}_E} \qquad (1)$$

where $\overline{D}_o$ is the observed average distance between each point and its nearest neighbours, $\overline{D}_E$ is the expected average distance between each point and its nearest neighbours.

$$\overline{D}_o = \frac{\sum_{i=1}^{n} d_i}{n} \qquad (2)$$

where $d_i$ is the distance between the point and its nearest neighbours, and $n$ represents the number of incident points.

$$\overline{D}_E = \frac{0.5}{\sqrt{n/A}} \qquad (3)$$

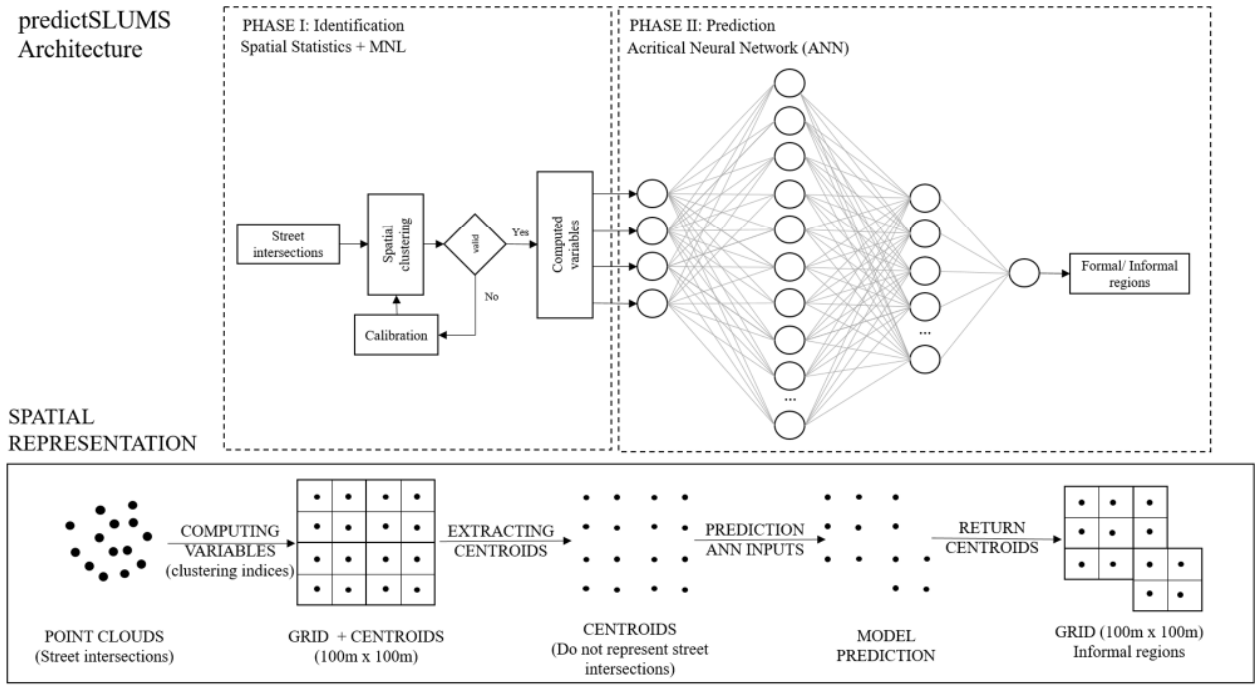

FIG. 2
PREDICTSLUMS MODEL

where $A$ is the area of the minimum enclosed rectangle for the incident points.

The null hypothesis is rejected, when the value of Nn is less than one, consequently, the intersection points are significantly clustered. However, this analysis only shows significances of clustering or disparity but does not interpret the type of clustering spatially nor the behaviour of the clustering over a range of different distances.

Second, we have computed Multi-Distance Spatial Clustering Analysis, knowing as Ripley's K-function (Ripley, 1981). This analysis aims to depict the change in the classification of the incident points, whether clustered or dispersed, over a range of band distances. The K-function, known as L(d) is calculated as:

$$L(d) = \sqrt{\frac{A\sum_{i=1}^{n}\sum_{j=1, j\neq i}^{n} k_{i,j}}{\pi n(n-1)}} \quad (4)$$

where $d$ represents the distance, $n$ represents the number of points, A represents the total area that confined all points, and $k_{i,j}$ represents the weight. For minimizing the computing time, the weight is set to 1 when the distance between i and j is less than d, as there is no edge correction.

As a result, at a given distance, the distribution of the incident points is considered to be more clustered than a random distribution when the k-observed value is larger than the k-expected one; reciprocally, the distribution is considered to be dispersed. In order for the k-value to be identified as a statistically significant cluster, the k-observed value must be larger than the upper confidence envelope[4].

The percolation theory can illustrate the hierarchy and the change in classification of random points in a given space within a range of distances, in terms of clustering size (Elsa Arcaute et al., 2016; Christensen, 2002; Piovani, Molinero, & Wilson, 2017). Yet, it does not explain how significantly dispersed the neighbouring points surrounding these clustered groups are within a given distance. Consequently, in order to identify the significant spatial variation of the points classification at a given distance, in term of low and high clustered or dispersed groups, we conducted an "optimised" Getis-Ord Gi analysis, known as a Hot Spot Analysis (Getis & Ord, 1992). This method aims to identify clusters of concentration of high or low values.

The index for Getis-Ord local statistics (Gi*) represents the GiZscore and it is calculated as:

$$G_i^* = \frac{\sum_{j=1}^{n} w_{i,j} x_j - \bar{X} \sum_{j=1}^{n} w_{i,j}}{S \sqrt{\frac{\left[n \sum_{j=1}^{n} w_{i,j}^2 - \left(\sum_{j=1}^{n} w_{i,j}\right)^2\right]}{n-1}}} \quad (5)$$

where $x_j$ is the value of point j, $w_{i,j}$ represents the spatial weight between point i and j, n is the number of incident points.

$$\bar{X} = \frac{\sum_{j=1}^{n} x_j}{n} \quad (6)$$

$$S = \sqrt{\frac{\sum_{j=1}^{n} x_j^2}{n} - (\bar{X})^2} \quad (7)$$

The larger the positive value of Gi*, the more intense the clustering of points of high values. When the results are statistically significant they are referred to as hot spots. For negative statistically significant values of Gi*, the smaller the value, the more intense the clustering of points of low values are. These are referred to as cold spots.

In order to reduce the critical thresholds of the p-value from a statistical point of view, first, GiZscore values are optimized based on a False Discovery Rate (FDR) correction method (Benjamini & Hochberg, 1995). This method is used to eliminate errors associated with spatial dependency and multiple testing. It is computed as the expectation of Q:

$$FDR(Q_e) = E(Q) = E\left(\frac{V}{V+S}\right) \quad (8)$$

Where v indicates the false positive rates, s indicates the true positive rates.

Second, while the model relies on GiZscore results, as a step forward for further optimization and for other cities of different context, Local Moran's I clustering and outlier method (Anselin, 1995) may be significant as it identifies the significance boundaries between points that can depict whether a neighbouring point belongs to an adjust group or not. This method can differentiate the classification of points, not only in terms of high and low values of clustering or dispersion but also from the local and global scale of all neighbouring points, resulting in identifying spatial outliers. However, this method is only a complementary one to the main GiZscore indices to ensure that the model can fit to different cities, bearing in mind the subtleties of the local setting of the variables that may vary from city to city, nonetheless, the administrative boundaries of cities that are not necessarily a cutting edge from the global spatial perspective.

*3.1.2 Calibration*

First, in order to compute GiZscore that define the optimum clustering of hot spots in cities, pinpointing the optimum distance (computed in Equation 4) requires manual calibration through trial and error before reaching the classification of the incidents points that is statistically significant within a band distance that is representative of the informal and slum areas in cities. After computing the statistically significant GiZscore where the hot spots are more likely to represent informal settlements, calibration for prediction is done automatically relying on the training sample of the machine learning algorithms that will be discussed in phase II.

*3.1.3 Cross-validation and prediction of hot-spots*

In order to cross-validate the outcome of the classification that represents informal settlements, two methods are used, one parametric and one non-parametric, in order to avoid the assumption of normal distribution of the data.

- *Method 1: Independent samples t-test analysis*

This method is used to compare the means of the GiZscore for both groups; informal, and formal regions. When the p-value is less than 0.05, the null hypothesis can be rejected and the results of GiZscore can be illustrated as statistically significant. It can be used to differentiate between formal and informal zones. For further explanation regarding t-test analysis, see Hoffman (2015); Smalheiser (2017).

---

[4] The confidence envelope is constructed based on a random distribution of the intersection points, so-called permutation, based on Monte Carlo test.

- *Method 2: Muli-Nominal Logistic (MNL) Regression model*

Unlike the previous method where GiZscore was used as a continuous numerical variable, in this method, it is defined as a discrete variable, whereas the statistically significant values of GiZscore, can represent three categories; Hot-spot, Not significant, and Cold-spot. For the sake of this study, we are interested in understanding how these categories of hotspot analysis are associated with the informality status and the numbers of neighbouring intersection points, in order to cross-validate the outcome of the hotspot classification with informal regions. Hence, a Multi-Nominal Logistic (MNL) regression model is conducted to analyze the collinearity among the three categories of GiZscore, as a dependent variable, in relation to the informality status, and the change in a number of neighbouring intersection points as two independent variables. For further explanation regarding discrete choice models and utility functions, see Ben-Akiva et al. (1997); Schroeder (2010).

The utility function of alternative hotspot category i in the occurrence of j is computed as:

$$v_{ij} = \varepsilon + \sum_{k \in T} b_k x_{ijk} \quad (9)$$

where $x_{ijk}$ represents the attribute k for point j on hotspot occurrence of i, $b_k$ is a coefficient in the utility function, $T$ represents the set of attributes, $\varepsilon$ represents the stochastic part of the utility function.

The coefficient of the MNL model is computed by estimating the maximum likelihood, whereas the stochastic part $\varepsilon$ is computed by assuming it as a double exponential distribution. The logarithm of the likelihood of the MNL model of the actual occurrence of hotspots can be expressed as:

$$Log\ L = \sum_{i=1}^{N} \sum_{j=1}^{J} Y_{ij} \ln P_i(Y = j/x, \beta),$$

where $P_i(Y = j/x, \beta) = \dfrac{\exp(v_{ij}(x_{ij,b}))}{\sum_{h=1}^{j} \exp((v_{ih}(x_{ih,b}))}$ (10)

where $Y$ is the dependent variable of the three categories of hot spots, X represents the independent variables, $v_{ij}$ is the utility function for jth alternative of ith choice, $N$ represents the occasion of choices, $j$ represents the number of alternatives, $P_i$ represents the predicted probability of the occasion of i category of hot spots, $\beta$ represents the parameter vector of the model,

In order to validate the fitness of the model, three measures were conducted; pseudo R-square, the significance of Chi-square, and the confusion matrix of the actual values and the predicted ones for each city.

### 3.2 Phase II: Prediction of informal regions and slums

Globally, not all informal settlements are well-defined and delineated in cities. Thus, it may add challenges for the cross-validation and self-calibration when computing the model. Hence, we introduce in this model a predicting method of unlabelled data, in which it is perceived as an indispensable feature for predictive modelling (Blum & Mitchell, 1998; Liang, Mukherjee, & West, 2007; Wu, Zhao, Qin, Lai, & Liu, 2017). We attempt to identify the informality in cities by training the model with well-defined and calibrated model of other cities with similar context, i.e. same region, or country.

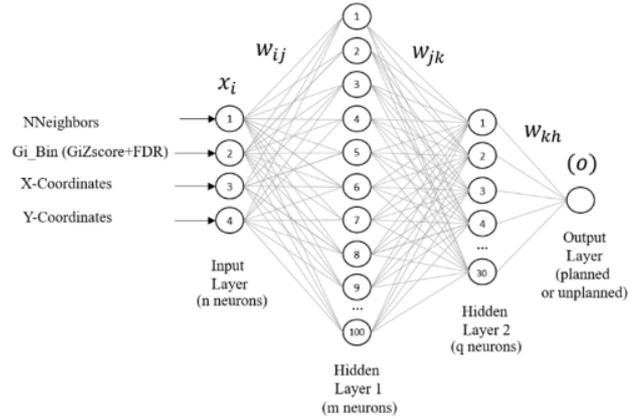

FIG. 3
PREDICTSLUMS-ANN ARCHITECTURE

### 3.2.1 ANN model architecture

We introduce Artificial Neural Networks (ANN) for classification and prediction. For the purpose of this study, unlike other machine learning algorithms for classification (i.e. Kernel Support Vector Machine), ANN model performs better with minimal input variables and with a large dataset for nonlinear classification tasks[5]. This makes ANN a good fit for the research purpose. Additionally, the accuracy of the model and the required time for computing it are another advantages. For further explanation, see Chen (1990); Goh (1995); Guan, Wang, & Clarke (2005); Nefeslioglu, Gokceoglu, & Sonmez (2008); Pijanowski, Pithadia, Shellito, & Alexandridis (2005); Pradhan, Lee, & Buchroithner (2010).

The ANN models are trained on the data sets of the studied cities in order to identify the current state of the informal settlements within the same city or elsewhere that can be used as a predictive model for identifying informal areas in a new data sets for the future of these cities or the surrounding context.

Fig. 3 shows the general architecture of the ANN model. Four input variables are used for ANN models that are previously computed in the first phase; GiZscores (three categorical variables), NNeighbor, the X and Y-Coordinates of the hot spot points. Based on trial and error, these coordinates enhance the performance of the model to better understand the profile of informal sprawl especially when the training and testing are performed for a single city, whereas the role of the coordinates is marginalized when the model is trained and tested for a group of cities.

All the input variables are covariates, whereas the dependent variable of the models (Y), is discrete nominal variable coded as 0 and 1 that represents informality. In order to operate with the algorithms of the ANN model, the input variables are standardized. After several trial and error for tuning the hyper-parameters of the ANN model, these input variables are fed-forward to two hidden layers of 100 and 30 neurons respectively, to output a single neuron that classifies the planning status of the region.

The training algorithm of the model is based on the back-propagation of error to update the weights of the neurons. It is compiled based on the optimization algorithm of a stochastic gradient descent with an initial learning rate of 0.001, relying on 'adam' optimizer (Kingma & Ba, 2014). The dataset for each city is divided randomly into training and testing (that is used for validation) sets in a portion of 70% and 30% respectively. The model is trained by feeding the training set as mini-batches of size 10, in which the weights of the neurons are adjusted after each mini-batch. This process of training and validation is repeated by 600 training cycles (epochs).

---

[5] Based on trial and model selection, the proposed ANN architecture has performed significantly better than a Kernel Support Vector Machine model for Greater Cairo with 6.4% increase in accuracy.

The output of the neurons for each layer of the ANN model is computed based on the general formula:
$$Y = s(\sum_{i}^{n} w_i x_i + b) \qquad (11)$$
where $n$ is the total number of input neurons, $x_i$ represents each input neuron, $w_i$ represents its weight, $b$ represents the bias, $s$ represents the activation function of the layer.

The input and the two hidden layers are activated based on Rectified Linear Unit (ReLU) to increase the non-linearity of the model and enhance the performance of the neurons (Dahl, Sainath, & Hinton, 2013; Glorot, Bordes, & Bengio, 2011). It is computed as ($s_1$):
$$s_1 = f(x) = \max(0, x) \qquad (12)$$
The output of the model is activated based on a sigmoid function ($s_2$). It is computed as:
$$s_2 = \delta(x) = \frac{1}{1+e^{-x}} \qquad (13)$$
The overall feed-forward propagation of the ANN model for the single output neuron can be expressed as:
$$Net_o = s_2 \left[ \sum_{k=1}^{q} w_{kh} * s_1 \left[ \sum_{j=1}^{m} w_{jk} \left( \sum_{i=1}^{n} w_{ij} x_i \right) \right] \right] \qquad (14)$$
where n is the total number of neurons in the input layer, m is the total number of neurons in the first hidden layer, q is the total number of neurons in the second hidden layer, $w_{ij}$ represents the weight between the neuron i in the input layer and the first hidden layer, $w_{jk}$ represents the weights of neuron j in the first hidden layer and the second hidden layers, $w_{kh}$ represents the weights between the neuron k in the second hidden layer and the output neuron.

*3.2.2 ANN model accuracy and cross-validation*

The accuracy of the ANN models is validated based on a Loss Function of Cross-Entropy error (E) method for n-class (Golik, Doetsch, & Ney, 2013; Janocha & Czarnecki, 2017). E is calculated as:
$$E = -\sum_{i}^{n} t_i \log(y_i) \qquad (15)$$
where $t_i$ is the target vector, $y_i$ is the output vector, n is the number of classes for classification (n=2).

In order to validate the performance of the model throughout the entire dataset, we have computed K-Fold Cross Validation for each model. Each dataset is divided into K-segmentations (K=10) where the model is computed 10 times. Each one is trained on (K-1) folds and validated in the remaining fold. This can represent the mean and the variance of the accuracies of the model through the different segments of the dataset (Jiang & Wang, 2017). For each value of K, the model is fitted with λ as an estimated parameter for the other k-1 folds to give $\hat{\beta}^{-k}(\lambda)$, whereas the K-Fold Cross Validation error is computed as:
$$CV(\lambda) = \frac{1}{K} \sum_{k=1}^{k} \sum_{i \in kth\ part} (y_i - x_i \hat{\beta}^{-k}(\lambda))^2 \qquad (16)$$
where $x_i$ is the independent variables of the model, $y_i$ the dependent variable.

Furthermore, we also computed the confusion matrix of the actual and predicted values to assess the false negative and positive of the model classification for the entire dataset.

*3.3 predictSLUMS model summary*

Fig. 4 shows the flowchart of the predictSLUMS model. It comprises the consecutive steps of the model, highlighting the two discussed phases and three main processes of the model; algorithms, calibration, and validation. The first section of the flowchart shows how street intersections are classified relying on a different set of spatial statistical analysis and different variables are computed. The second section shows the manual calibration and validation processes of the model, whereas the last section shows how the proposed model can predict informality in the same city, or elsewhere by training the model in cities with a valid classification. This feature can allow the model to function in cities that not necessary holds official data for validation.

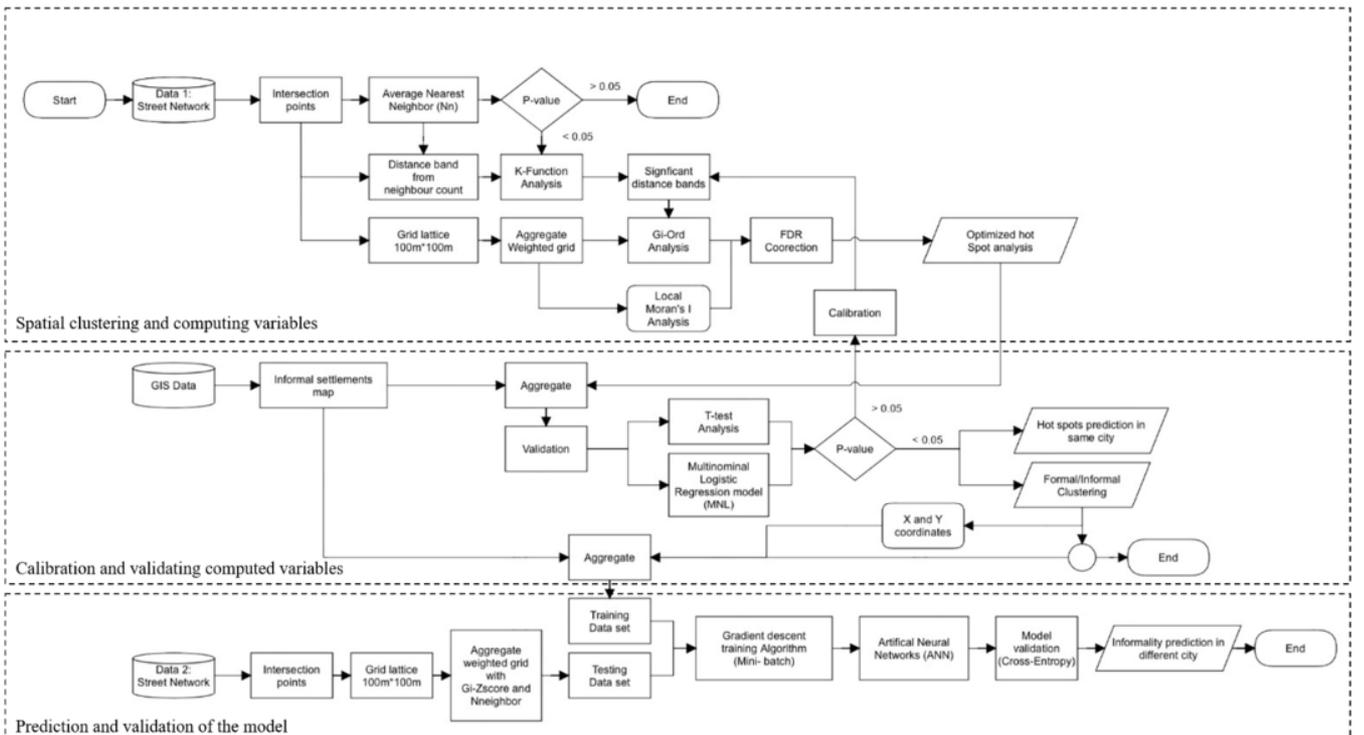

FIG. 4
PREDICTSLUMS FLOWCHART

## 4. CONTEXT AND CASE STUDIES

We have applied the predictSLUMS model for five cities; four Egyptian cities, and one Indian city. While we focus on the Middle East and North Africa (MENA) context, we also aim to verify the algorithms of the model globally without being confined to a certain context.

### 4.1 Egyptian case studies

Egypt, one of the most populated countries in the MENA region, has the biggest portion of informal settlements in the region which are locally-known as Ashwa'iat (Ibrahim & Masoumi, 2018; Sims, 2013; UN-Habitat, 2012). Out of a total population of approx. 93 million, at least 40 million live in informal housing. This makes Egyptian cities a crucial case for identifying and predicting informal settlements. Less than 5% of the Ashwa'iat in Egypt can be considered as unsafe areas or slums. The model deals with all types of Ashwai'at and slums without classification of the types or the degree of severity of the informal areas.

We have computed the model for four Egyptian cities with diverse urban sprawl profiles; Greater Cairo (including Cairo, Giza, Qaliobya cities, and two new towns; 6$^{th}$ of October and New Cairo), Alexandria, Hurghada, and Minya. There were five main criteria for selecting these case studies; the size and population of the city, geographical location, rural-urban structure, and forms of informal area (See Fig. 5).

This paper focuses on representing the model results in the case of Greater Cairo in detail, however, the model results for the other cities will be discussed briefly.

There are two reasons that Greater Cairo is an appropriate case study for identifying and predicting housing informality. Firstly, at least 70% of urban dwellers live in informal housing. Secondly, there is a diversity of informal sprawl and a wide spectrum of urban patterns.

The city of Alexandria is the second biggest urban agglomeration in Egypt, in which at least 40% of housing is informal. Due to the spatial configuration and constraints of the city, the informal areas are rather more compacted and have grown more vertically in comparison to Greater Cairo. Unlike other cities in Egypt, informal housing there can take the form of a high-rise building of more than ten floors.

The city of Minya is located in upper Egypt. It is characterized by rural development. Informal sprawl mainly takes place in rural areas in the form of relatively dispersed low-rise buildings.

Finally, the city of Hurghada is located by the Red Sea where the development of the city is based on the tourism sector. The informal areas are mainly characterised by low-rise scattered housing in desert hinterland and core areas of the city.

### 4.1 Indian case study

To verify the predictSLUMS model in another context, we have computed the model for Mumbai in India. What makes the context of Mumbai different from the ones in Egypt is that slums there have various densities and forms. This adds a new dimension for identifying slums and informal areas in cities of higher densities, and demonstrates the possibility of using this model in the various contexts of cities of the global south.

### 4.2 Data

For the Egyptian case studies, we used the official GIS data surveyed by the General Organization of Physical Planning in Egypt (GOPP) for the four Egyptian cities (GOPP, 2014).

For the case of Mumbai, we used a raster image that represents the official the planning status to calibrate and validate the model, whereas, for the spatial network, we have used the open source data provided by OpenStreetMap (OSM) as input data for classification and prediction.

#### 4.2.1 Street network data

The street network data of Greater Cairo comprises 194,869 street intersection points (nodes), including endpoints and a total length of 18,636.5 km. Data of Alexandria contains 63,441 nodes and of a total street length of 5,647.7 km. Data of Hurghada includes 9,611 nodes and of a total street length of 1,058.8 km. last, Minya contains 4,604 nodes and the total street length is 329.6 km (See Fig. 6). What makes this data a better dataset than that provided by OSM is that the local streets of both formal and informal areas are well-delineated by field surveys that contains the entire hierarchy of street network, including the local ones in the informal areas and the delineated slums. This enhances the precision of the model when relying on the densities of street intersections.

For the case of Mumbai, as we aim to introduce the concept of the predictSLUMS model regardless to the type of data that may or may not be available elsewhere. The OSM data comprises 30,110 nodes, and of a total length of 4,182.8 km.

We have used the street intersections data from the original form of the spatial networks for all case studies without performing any filtrations or data processing.

#### 4.2.2 Informality status data

To validate the model results for the case of the Egyptian cities, we used the data of the informal regions defined by GOPP for the four studied cities. The planning status is defined either formal or informal, including unplanned areas and slums, and is mainly based on official field surveys. This classification is based on four main criteria – planning zones, land ownership, building standards and location. Land that is officially planned, is developed by acquiring a building licence, and the building follows the area code of the neighbourhood is considered to be formally developed. Land that is owned properly but the housing is not built properly or officially is considered informally developed. The method of land acquisition and ownership is another crucial issue; informal areas include those built on state land, or land in other forms of ownership, that does not give individuals the right to build their houses. Housing built on land not identified through the official land-use plan of the city as being available for this purpose is also considered informal. For example, housing that is built on agriculture land. Finally, housing units that are built on the periphery or the hinterland of the administrative urban boundary of the city is considered informal housing. There are a wide range of literature that review informality of housing in

Cairo and other Egyptian cities, in which it is in line with the official census used in this study (Harris & Wahba, 2002; Hassan, 2012; M. Ibrahim, 2017; O'Donnell, 2010; Sims, 2010, 2015; Sims, Sejoume, & El Shorbagi, 2003; Soliman, 2007, 2012).

To validate the results for the case of Mumbai, we have used a map that identifies the formal and the informal regions in Mumbai that is open-access and provided by PK Das & Associates (2011). The map is prepared based on google maps and the official development plan from Municipal Corporation of Greater Mumbai in 2011. The output of this map is in line with several studies related to slums and informal regions conducted in Mumbai (Bürgmann, 2015; Y. Zhang, 2017).

In order to operate with the model, we have processed this image by classifying the two categories of formal and informal areas relying on supervised image classification. The result is standardized for the model to a raster of grid cells of size 100m (See Fig. 7).

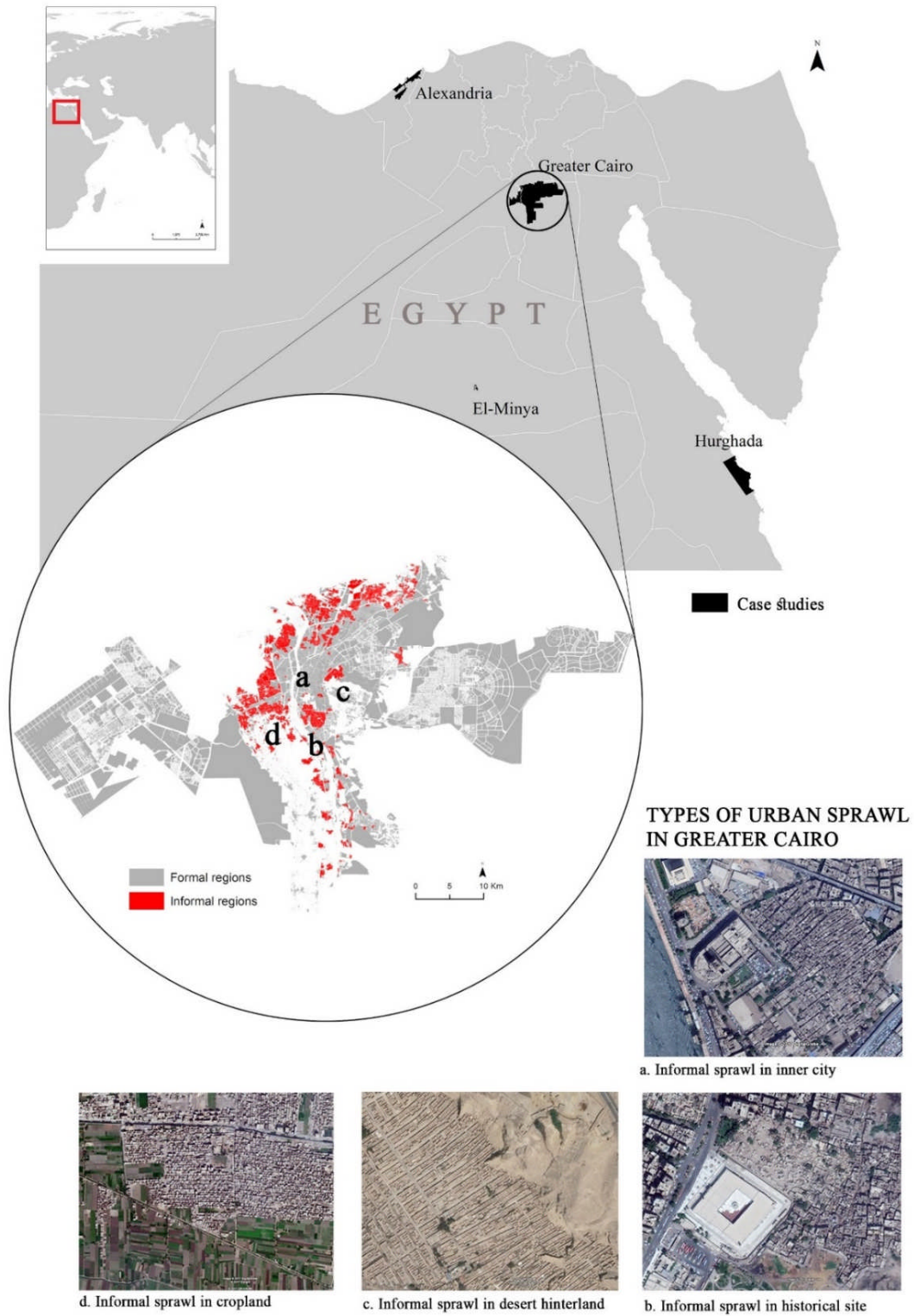

FIG. 5
EGYPTIAN CASE STUDIES AND HOUSING INFORMALITY IN GREATER CAIRO

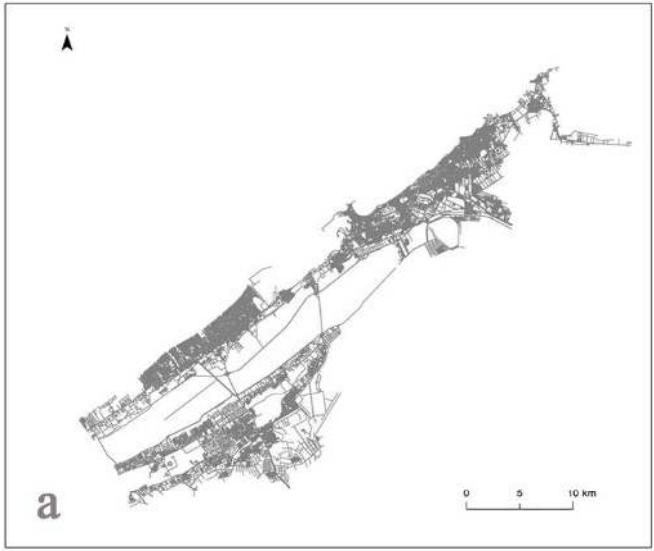
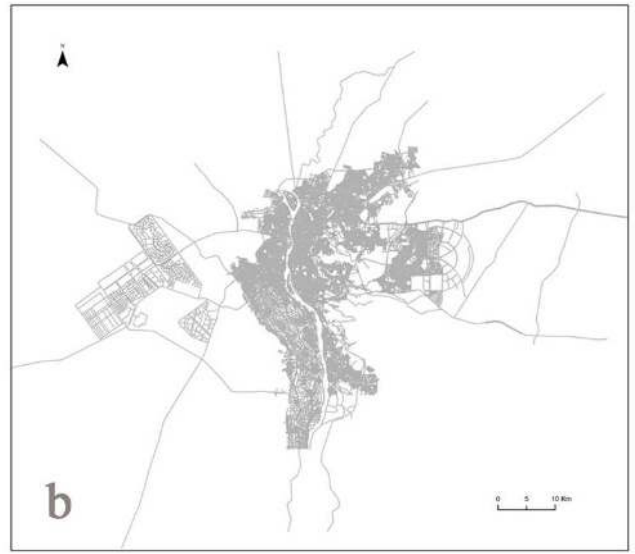
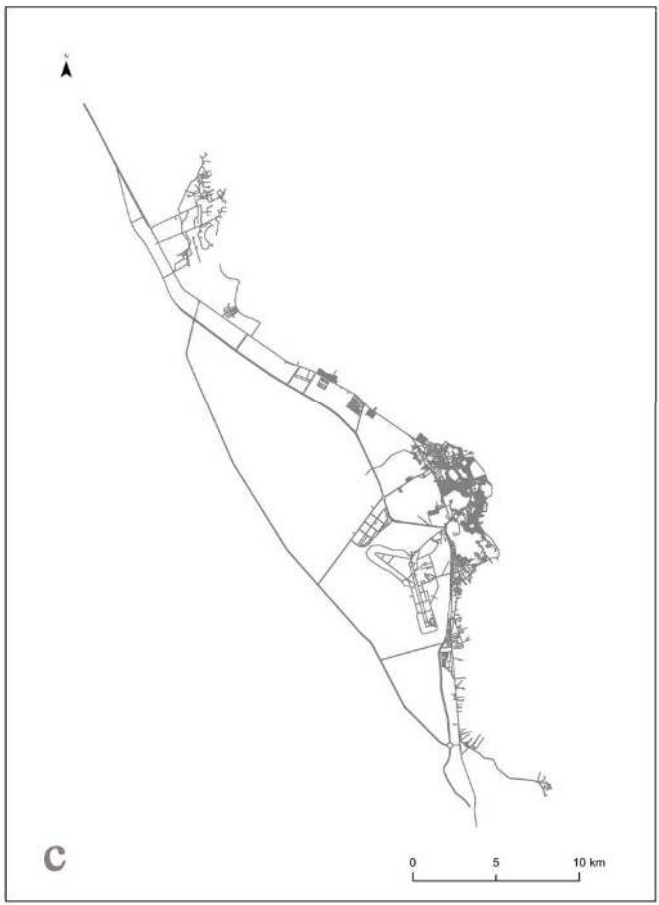
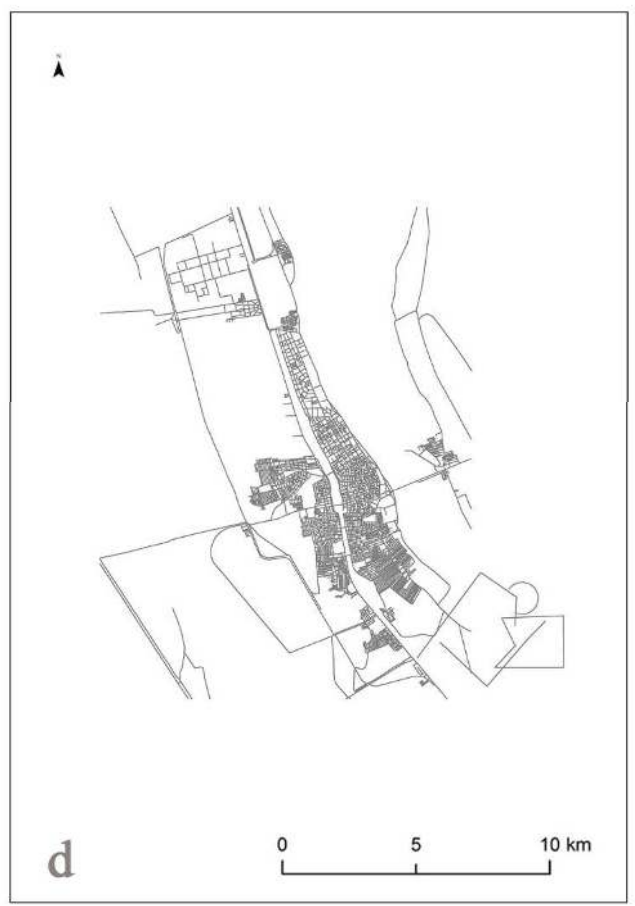

Fig. 6
STREET NETWORK DATA FOR THE EGYPTIAN CASE STUDIES
A) ALEXANDRIA CITY    B) GREATER CAIRO    C) HURGHADA CITY    D) MINYA CITY

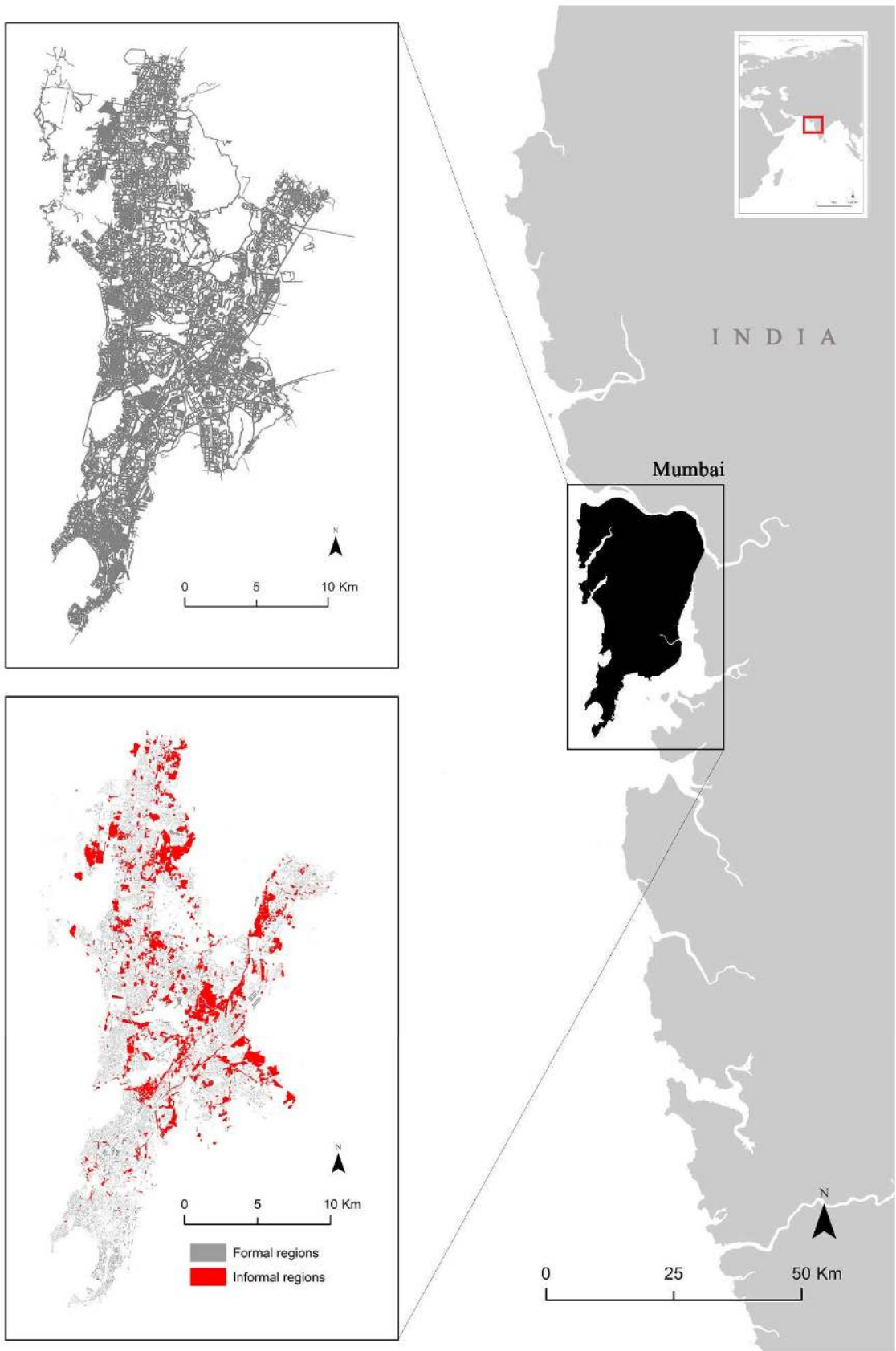

Fig. 7
Spatial Network and informality in Mumbai, india

## 5. RESULTS

### 5.1 Clustering indices of Greater Cairo

Computing Nn using Euclidean distance, gives a ratio between the observed and the expected average distance between incident points equal to 0.2, with a z-score equal to -666.3 (See Table 1). This is a statistically significant result, which means that there is a less than 1% likelihood that the pattern of the incident points is the result of a random distribution and suggests that the incident points are clustered.

TABLE 1
RESULTS OF AVERAGE NEAREST NEIGHBOURS ANALYSIS

| Observed Mean Distance | Expected Mean Distance | Nearest Neighbour Ratio (Nn) | z-score |
|---|---|---|---|
| 24.9133m | 118.1095m | 0.210934*** | -666.37019 |

The results of the K-function analysis conducted in order to understand the change in classification of points within an interval of distances, show that the points are significantly

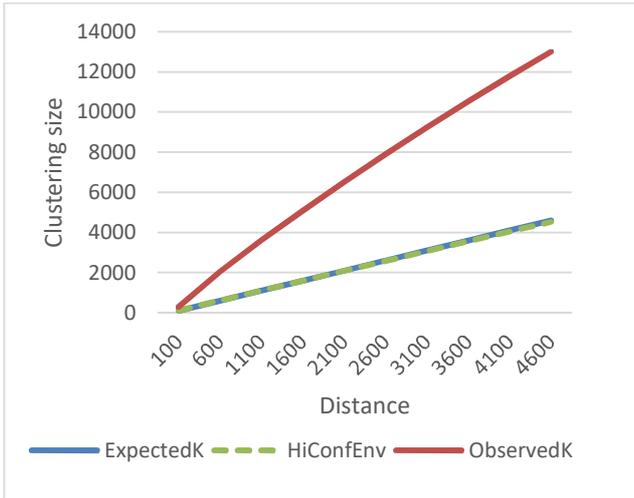
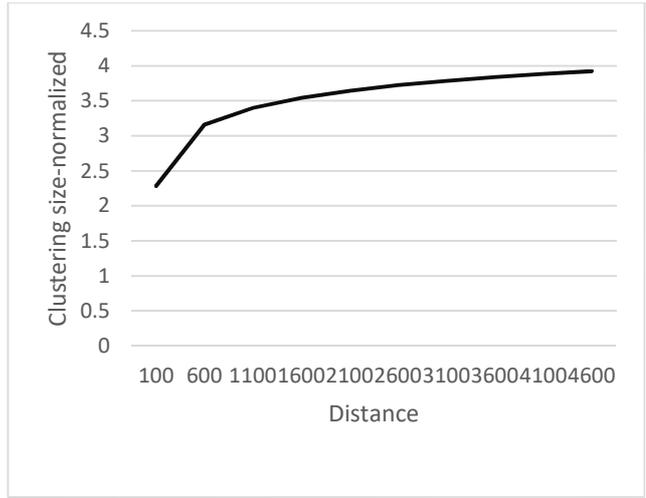

FIG. 8
A) RESULTS OF EXPECTED AND OBSERVED K-FUNCTION ANALYSIS OF GREATER CAIRO
B) RESULTS OF THE LOG DIFFERENCES OF OBSERVED AND EXPECTED VALUES OF GREATER CAIRO

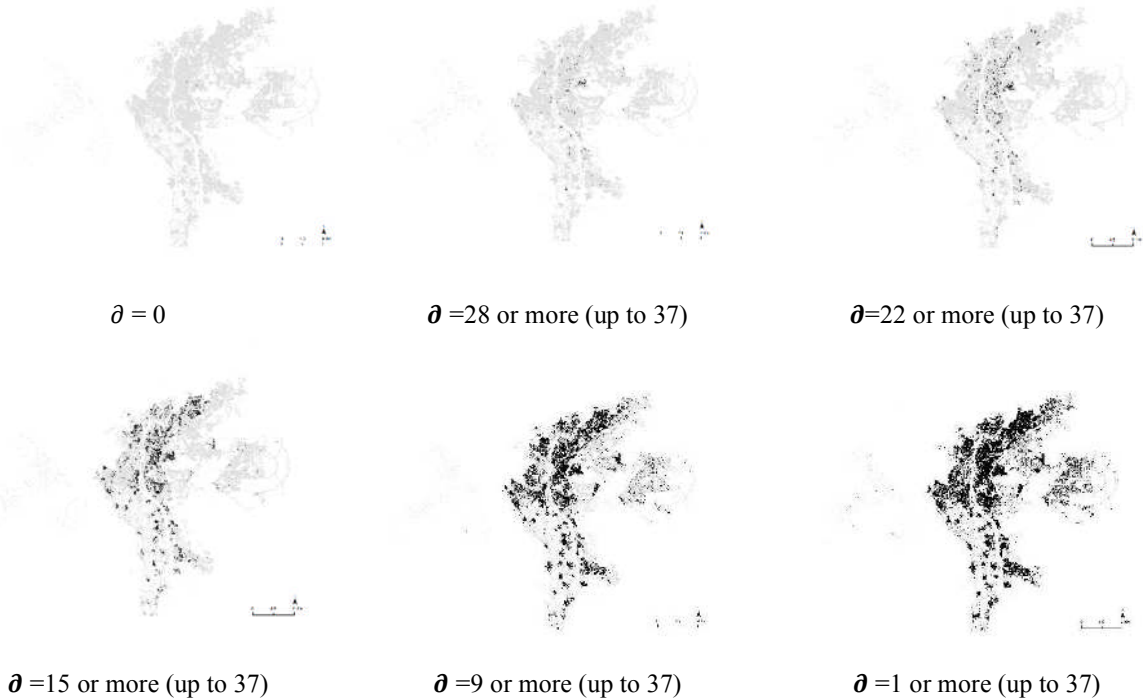

∂ = 0     ∂ = 28 or more (up to 37)     ∂ = 22 or more (up to 37)

∂ = 15 or more (up to 37)     ∂ = 9 or more (up to 37)     ∂ = 1 or more (up to 37)

FIG. 9
THE DENSITY OF STREET INTERSECTION OF GREATER CAIRO IN A LATTICE (100M x100M)

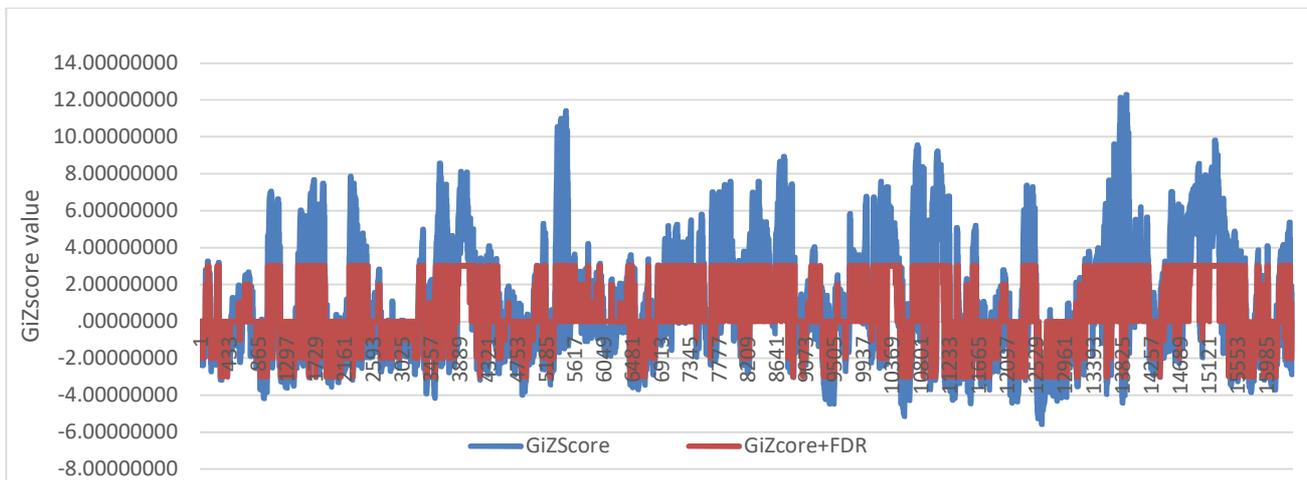

*Fig. 10*
*The relation between the results of GiZscore with and without FDR correction for Greater Cairo*

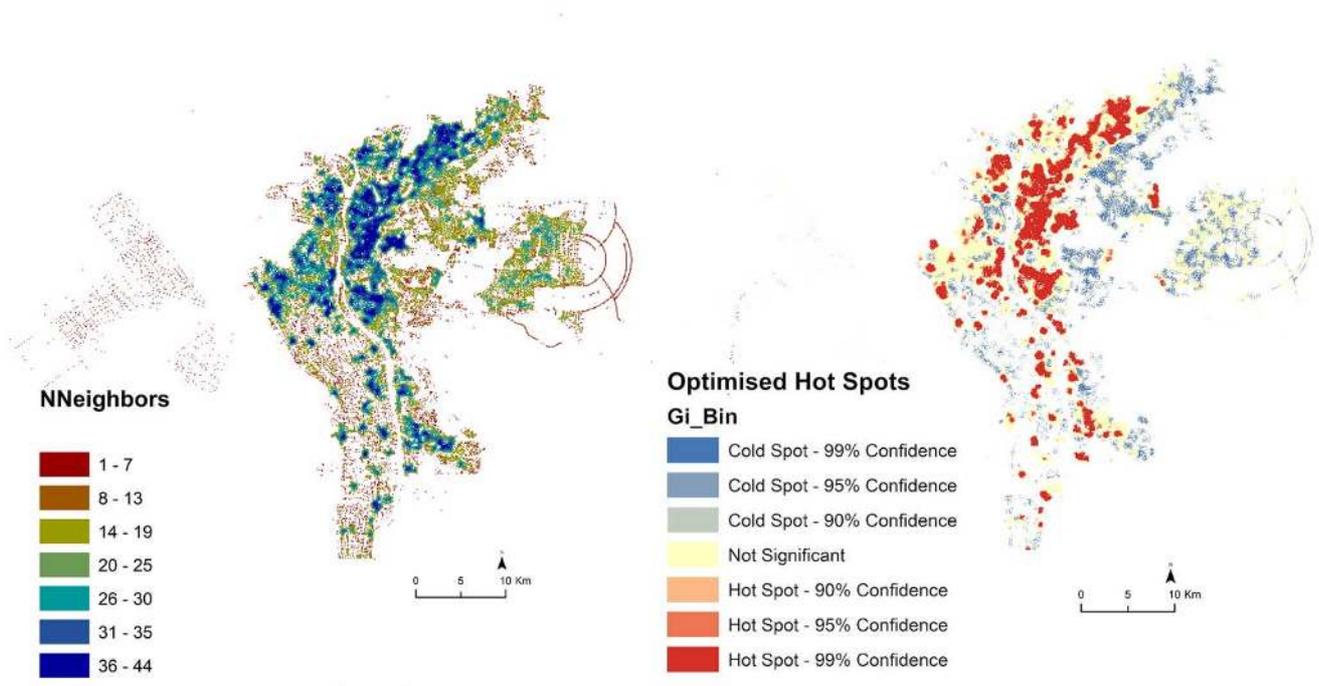

*Fig. 11*
*Statistically significant nearst neighbors and hot spots in Greater Cairo, at distance:334m (grid cell: 100m x 100m)*

clustered over the distance between 100m to 4600m since the observed k indices are larger than the expected k –value for each band distance, are greater than the high confidence envelope. The k-function suggests a noticeable change in the spatial change of classification of points at an approximate distance of 500m (See Fig. 8).

Fig. 9 illustrates selected cases that represent various clusters of Greater Cairo's neighbourhoods according to the density of street intersections. Based on visual inspection, the results show that the points are intensely clustered where informal settlements are located, beside the historical part of Greater Cairo within a higher intersections density in the core central region of the city. In contrast, the newer developed areas, i.e. New Cairo city (the area to the East) or 6th of October city (the area to the West), seem to be dispersed when compared to the older parts of Greater Cairo. However, these findings do not represent clusters that are statistically significant to informal regions.

TABLE 2
T-TEST STATISTICS AND MODEL PARAMETER

| Informality | | N | Mean | Std. Deviation | Std. Error Mean | t | df | Std. Error Difference | 95% Confidence Interval of the Difference | |
|---|---|---|---|---|---|---|---|---|---|---|
| | | | | | | | | | Lower | Upper |
| NNeighbors | Formal Area | 27976 | 46.192 | 21.982 | 0.131 | -88.420*** | 43,728.000 | 0.203 | -18.308 | -17.514 |
| | Informal Area | 15754 | 64.102 | 17.026 | 0.136 | -94.832*** | 39,572.944 | 0.189 | -18.281 | -17.541 |
| GiZScore | Formal Area | 27976 | 0.197 | 3.302 | 0.020 | -93.764*** | 43,728.000 | 0.032 | -3.100 | -2.973 |
| | Informal Area | 15754 | 3.234 | 3.158 | 0.025 | -94.933*** | 33,881.348 | 0.032 | -3.099 | -2.973 |

*** statistically significant at p-value equal to 0.000

TABLE 3
MNL PARAMETER ESTIMATES

| Hot and Cold Spots | | B | Std. Error | Wald | df | Exp(B) | 95% Confidence Interval for Exp(B) | |
|---|---|---|---|---|---|---|---|---|
| | | | | | | | Lower Bound | Upper Bound |
| Cold spot | Intercept | -0.421*** | 0.064 | 42.910 | 1.000 | | | |
| | NNeighbors | -0.041*** | 0.001 | 2,000.496 | 1.000 | 0.960 | 0.958 | 0.962 |
| | Formal Area | 1.603*** | 0.052 | 964.303 | 1.000 | 4.970 | 4.491 | 5.499 |
| | Informal Area | 0[b] | | | 0.000 | | | |
| Hot spot | Intercept | -4.657*** | 0.060 | 5,991.541 | 1.000 | | | |
| | NNeighbors | 0.085*** | 0.001 | 8,144.955 | 1.000 | 1.089 | 1.087 | 1.091 |
| | Formal Area | -0.626*** | 0.027 | 526.132 | 1.000 | 0.535 | 0.507 | 0.564 |
| | Informal Area | 0[b] | | | 0.000 | | | |

a. The reference category is: Not Significant.
b. This parameter is set to zero because it is redundant.
*** statistically significant at p-value equal to 0.000
Model Fitting: Likelihood Ratio test: Chi-Square equal to 28576.4 at p-value equal to 0.000

## 5.2 Optimized Hotspot analysis (statistically significant classification)

After applying the FDR correction method as shown in Fig. 10, the results of the Gi-Ord analysis illustrate that the hot and cold spot areas in Cairo are optimized where the threshold values are minimized for both positive and negative values.

At an optimized distance of 344m, GiZscores that represent the three categories (hot, not significant and cold spots) and the computed nearest neighbours in Greater Cairo show a high degree of representative results of informal regions when compared visually to the official data of informal settlements. There are fewer regions in New Cairo town that are labelled as hot spots that do not represent informal zones (See Fig. 11).

## 5.3 Model fit for Greater Cairo

### 5.3.1 Comparing the clustering indices with the status of the built environment (t-test)

Table 2 illustrates the results of the t-test analysis, using a 95% confidence level, for the two computed variables. The p-values for both predictors (GiZscores, and NNeighbors) are statistically significant. The changes in the values of these two variables are associated with the type of group, formal or informal, to which each incident point belongs. In the case of the group of informal areas, the mean number of neighbours is larger than for those in formal areas. Also, the mean of GiZscore is larger for those in informal areas than those in formal areas. This means that informal regions are more likely to be a hotspot with a denser number of intersection points than those in formal areas.

### 5.3.2 The association between hot spots and informal regions

In order to understand the association between hotspots, informal areas and the number of neighbouring intersection points, an MNL model was created (see Table 3 for the MNL model results for Greater Cairo). In general, the model shows statistically significant results for the two studied predictors, in relation to GiZscore as an independent variable with a good fit in term of significant chi-square value and likelihood ratio.

In the case of cold spots, the number of neighbouring intersection points is statistically less when compared to the reference category of not significant zones (negative B-value for NNeighbors). Also, cold spots tend to comprise of more formal zones than of informal ones when compared to the not significant areas (positive B-value for the formal area).

In the case of hot spots, the number of neighbouring intersections are likely to be denser when compared to the not significant zones (positive B-value for NNeighbors). On the other hand, hot spots tend to contain more informal areas than those regions categorized as not significant (negative B-value for the formal area).

Fig. 12 shows the residuals of the MNL model for the five studied cities (maximum iteration = 100). By looking at the table of predicted and observed values, the models show a good fit for the association and the prediction of hot spots in the cases of Cairo and Alexandria. The accuracy of prediction declines in the remaining three cases. However, in general, the model shows good validation for the correlation between the computed variables of the GiZscore and the NNeighbors with the informality data. This demonstrates that the ANN models can predict actual informality from such variables as those used here.

*5.4 ANN Models prediction and validation*

We have computed eight ANN models to provide evidence of predictSLUMS's reliability. The goal was not just to show the high accuracy of the model but to show the versatility of the predictSLUMS model, which can be trained in one city to predict the formal and the informal areas within the same city or another city. Table 4 gives a summary of the computed ANN models

Fig. 13 to Fig. 17 represent, in detail, the overall simulation of the predictSLUMS model. For each studied city, an ANN model has been computed to predict the informal regions there. The accuracies for both training and validation in regard to each training cycle (epoch) are illustrated. In order to assess the variance of the training dataset, the results of K-Fold Cross-Validation is shown. To further assess the performance of the model, a confusion matrix has been computed for the actual and predicted values for each category (formal/informal) for the entire dataset using the pre-trained models for each city, whereas an overall validation accuracy of the entire dataset is also illustrated.

Critical aspects of the training and validation of models are discussed below.

In the case of Greater Cairo, the average accuracy of the cross-validation is 87%, and the variance is 0.0087. The model shows a good accuracy that is nearly normalized over the different segments of data with an overall accuracy for the pre-trained model of 89.3%. The training and validation accuracies according to each training epoch show a uniform growth, highlighting the good fit of the training algorithm.

In the case of Alexandria, the average accuracy obtained from the K-fold cross-validation dropped to 83%, whereas the variance is 0.0097. The overall validation accuracy of the model is 83.6%

In the case of Hurghada, the average accuracy of the model is 91% and the variance is 0.027%. Unlike the cases of Greater Cairo and Alexandria, the deviation of the accuracies has increased, whereas the high threshold is 96% and the low threshold is 86%. However, the model shows an overall good classification accuracy for both formal and informal regions with a validation accuracy of pre-trained model of 93.3%.

In the case of Minya, the average accuracy of the cross-validation reached 98%. This is not due to an overfitting of the model; (this can be seen from the training and validation accuracy charts), but rather the small size and the simplicity of the city in comparison to the other studied cities.

In the case of Mumbai, the model shows the lowest average accuracy of cross-validation, which dropped to 80% with a good variance of 0.091. While the model shows a good fit of an overall validation accuracy in classifying informal regions and slums in Mumbai of 82.1%, there are several reasons that can explain such a drop in the accuracies. Most importantly, the spatial data used for analysis and the validation data used for training and calibration play an important role in such a drop. We used OSM data, which does not fully delineate all informal regions to the local street level; this has affected the results of the GiZscore values to be normalized among the different areas of the city. Accordingly, this has influenced the model performance.

Additionally, we have computed a single model for the four Egyptian cities, where it has been trained and tested on a random sample of their data, 70% and 30% respectively. The overall validation accuracy of this model is 81.1%. Similarly, we have trained a single model for the five cities, with a validation accuracy of 78.3%. Last, we have trained and tested a model in the data set of Cairo, Alexandria, Hurghada and Mumbai with a validation accuracy of 74.9%. We have used pre-trained model to predict the formal and the informal regions of Minya city, in which the model shows an accuracy of 71.2%. This model exemplifies the benefits of our novel approach for identifying and predicting slums by showing a single model can fit different cities regardless to their local context.

TABLE 4
NEURAL NETWORK MODELS SUMMARY

| ANN Models | | Sample random distribution | | N (valid) | Percent | Model overall validation accuracy (pre-trained model) | K-Fold Cross-Validation | |
|---|---|---|---|---|---|---|---|---|
| | | | | | | | 10-folds average accuracy | Variance |
| Prediction within the same city | Model 1 | Training | Greater Cairo | 30611 | 70% | 89.3% | 87% | 0.008 |
| | | Testing | | 13119 | 30% | | | |
| | Model 2 | Training | Alexandria | 16298 | 70% | 83.6% | 83% | 0.009 |
| | | Testing | | 6986 | 30% | | | |
| | Model 3 | Training | Hurghada | 1470 | 70% | 93.3% | 91% | 0.027 |
| | | Testing | | 631 | 30% | | | |
| | Model 4 | Training | Minya | 870 | 70% | 97.6% | 98% | 0.021 |
| | | Testing | | 373 | 30% | | | |
| | Model 5 | Training | Mumbai | 5227 | 70% | 82.1% | 80% | 0.019 |
| | | Testing | | 2241 | 30% | | | |
| Prediction in different cities | Model 6 | Training | The four Egyptian cities | 49250 | 70% | 81.1% | - | - |
| | | Testing | | 21108 | 30% | | | |
| | Model 7 | Training | The five case studies | 54478 | 70% | 78.3% | - | - |
| | | Testing | | 23348 | 30% | | | |
| | Model 8 | Training | Cairo, Alexandria, Hurghada and Mumbai | 53608 | 70% | 74.9% | - | - |
| | | Testing | | 22975 | 30% | | | |
| | | Prediction | Minya | 1243 | - | 71.2% | | |

In general, the training and validating accuracies of the ANN models in each city tend to be uniform. We have only applied features dropout regulation (0.5) after each hidden layer (Dahl et al., 2013; Srivastava, Hinton, Krizhevsky, Sutskever, & Salakhutdinov, 2014) for only model 8 to avoid overfitting when predicting formal and informal areas in Minya. This is may be relevant for future model when training the model in other cities.

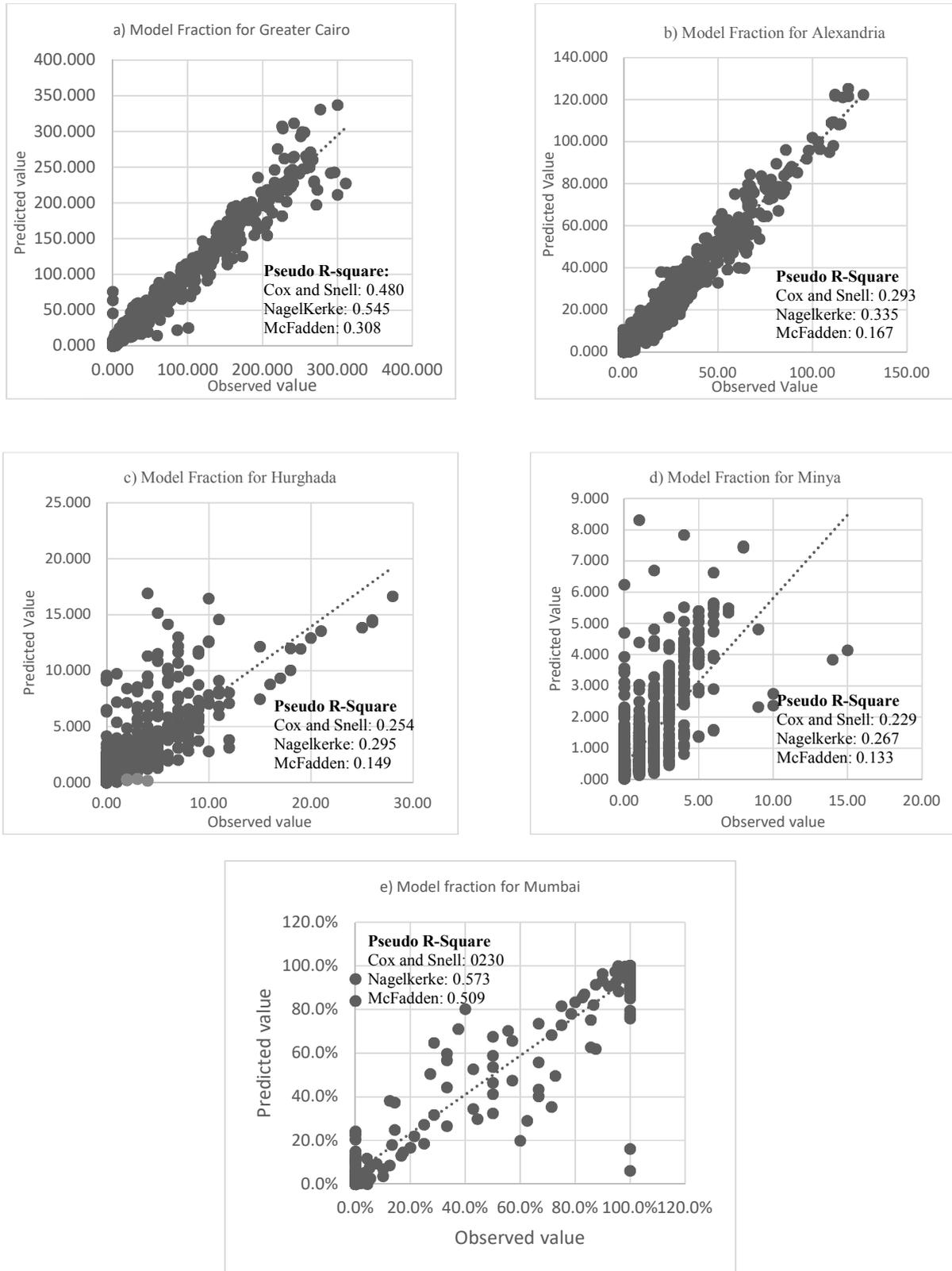

FIG. 12
MNL MODELS RESIDUALS FOR CASE STUDIES

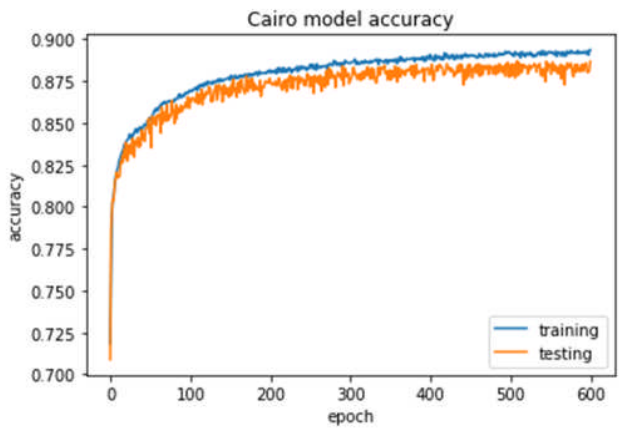
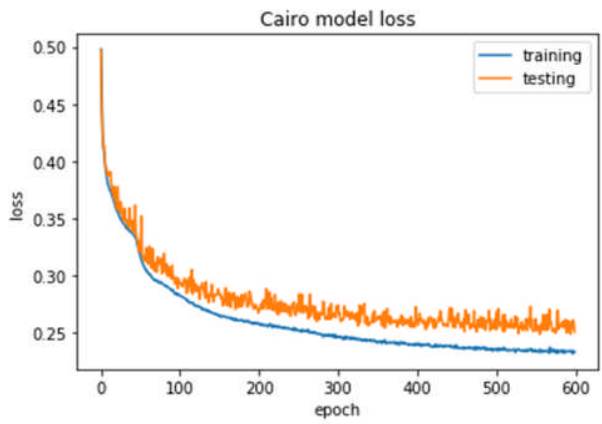
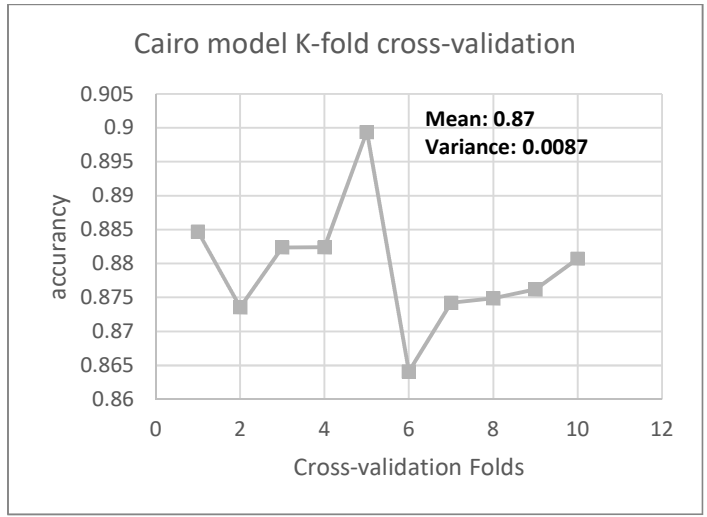

| Confusion matrix | | Actual values | |
|---|---|---|---|
| | | Formal | Informal |
| Predicted | Formal | 24690 | 3286 |
| | Informal | 1370 | 14384 |

Total cells: 43730
Overall validation accuracy: 89.3%

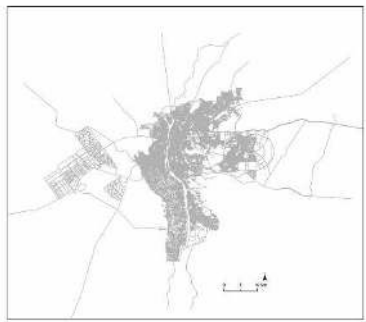
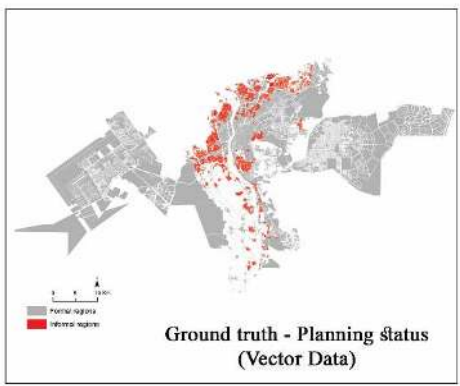
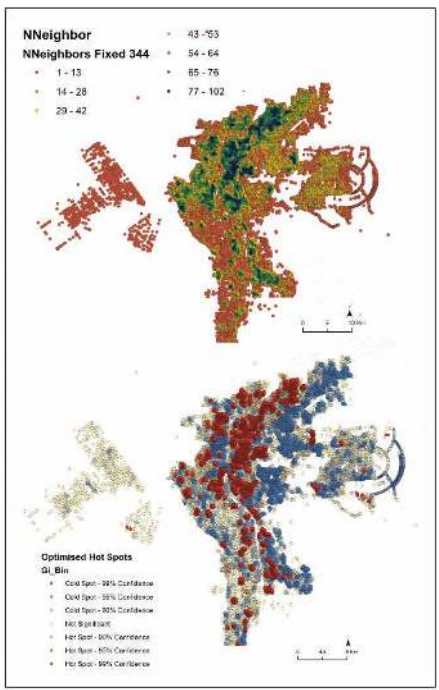
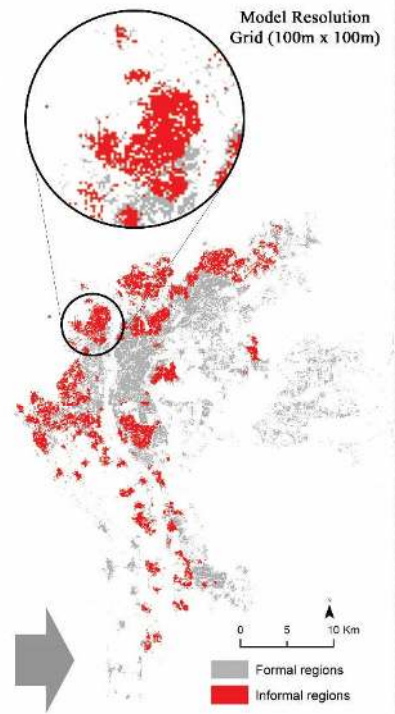

FIG. 13
predictSLUMS - Greater Cairo: identification, training and prediction

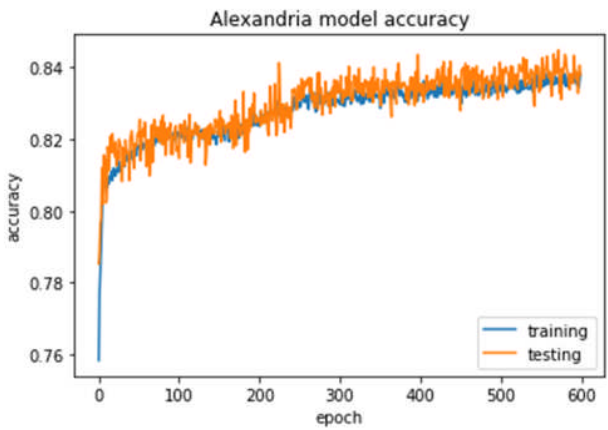
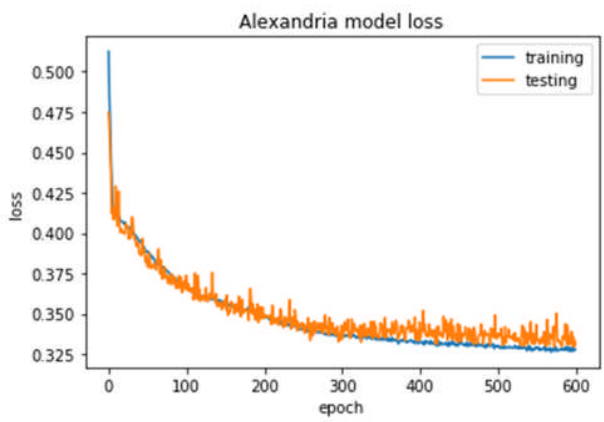
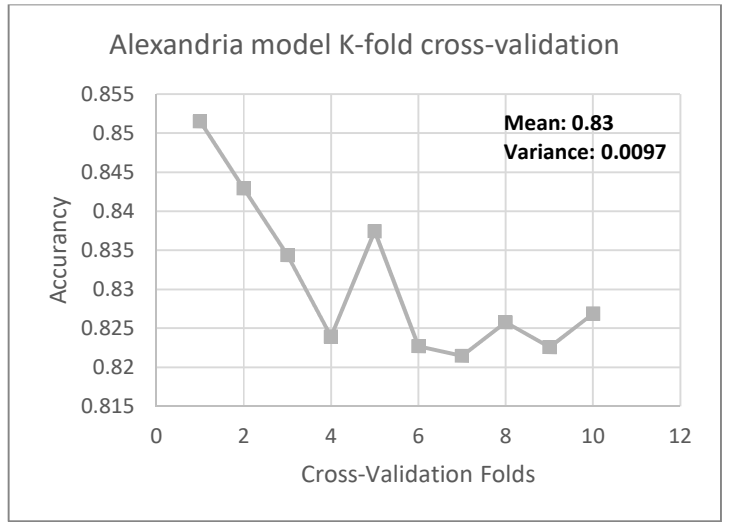

| Confusion matrix | | Actual values | |
| --- | --- | --- | --- |
| | | Formal | Informal |
| Predicted | Formal | 14910 | 1707 |
| | Informal | 2090 | 4577 |

Total cells: 23284
Overall validation accuracy: 83.6%

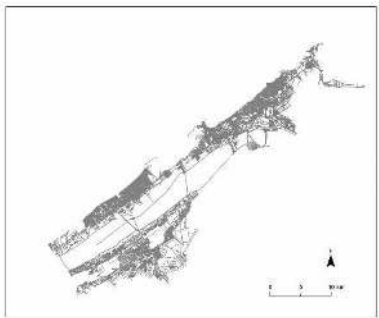
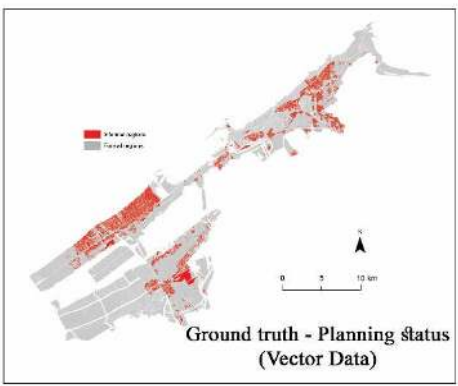
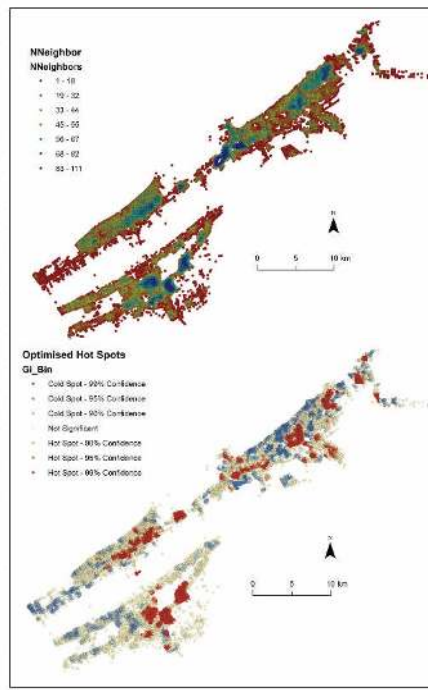
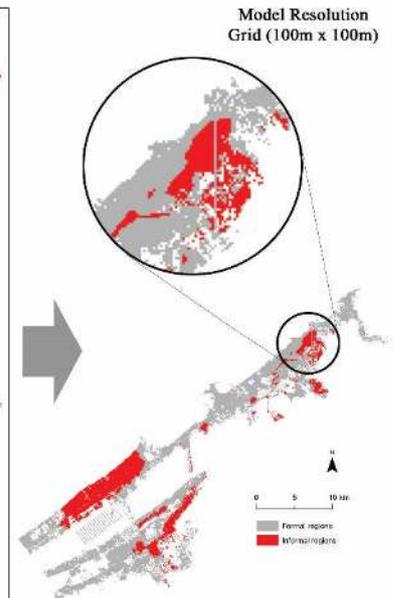

FIG. 14
predictSLUMS - Greater Cairo: identification, training and prediction

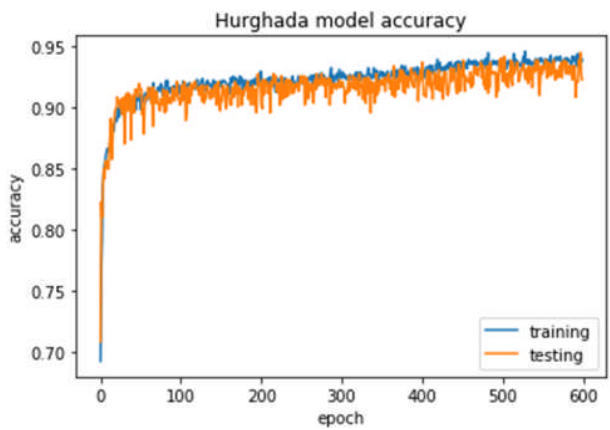
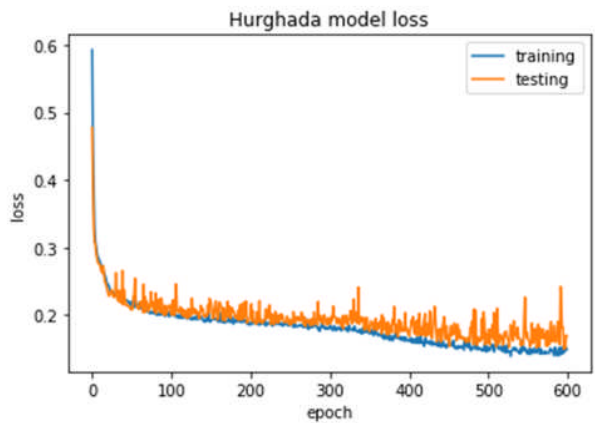
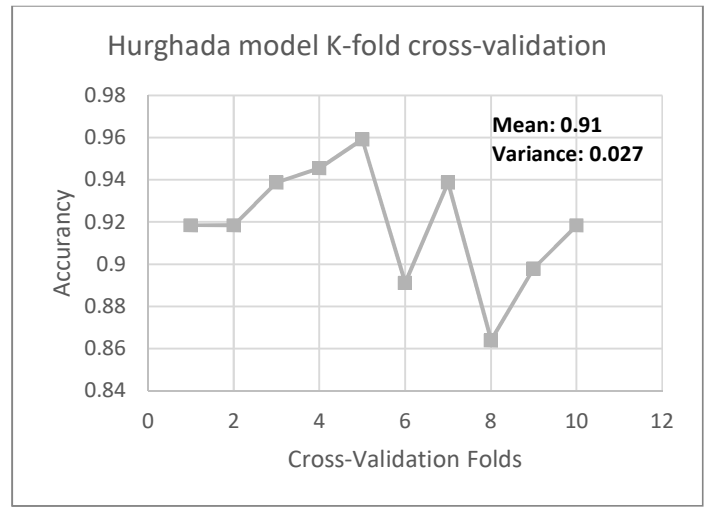

| Confusion matrix | | Actual values | |
|---|---|---|---|
| | | Formal | Informal |
| Predicted | Formal | 1372 | 92 |
| | Informal | 48 | 589 |

Total cells: 2101
Overall validation accuracy: 93.3%

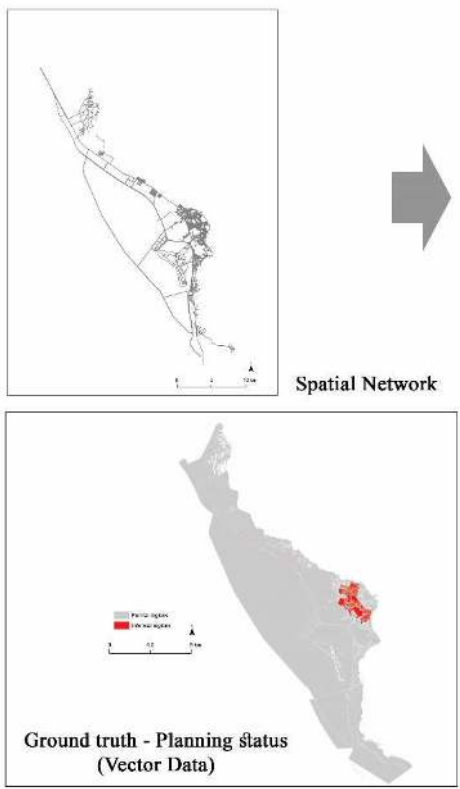
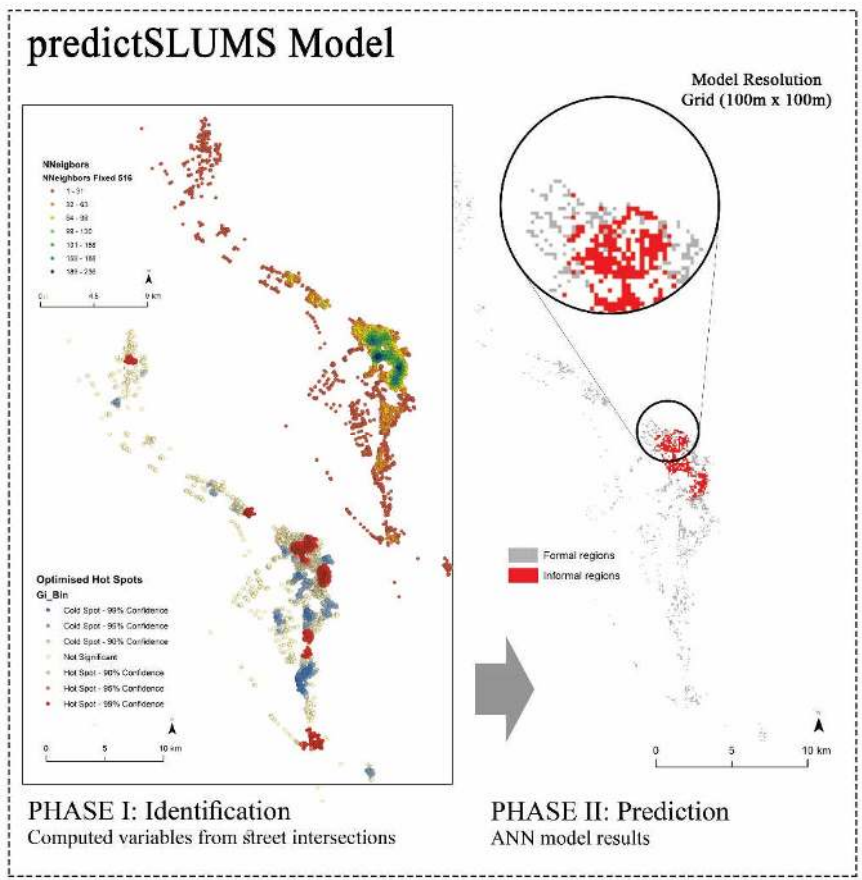

*FIG. 15*
*predictSLUMS - Hurghada: identification, training and prediction*

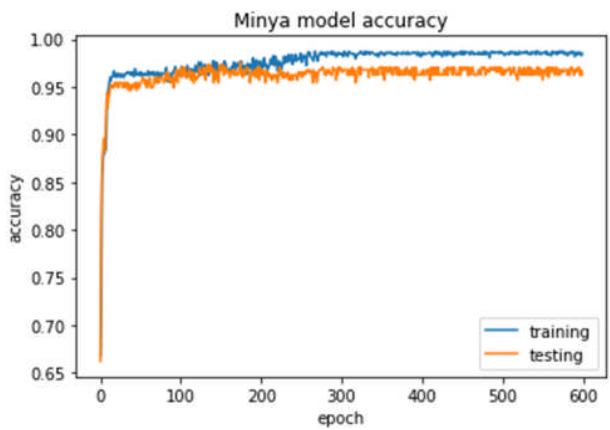

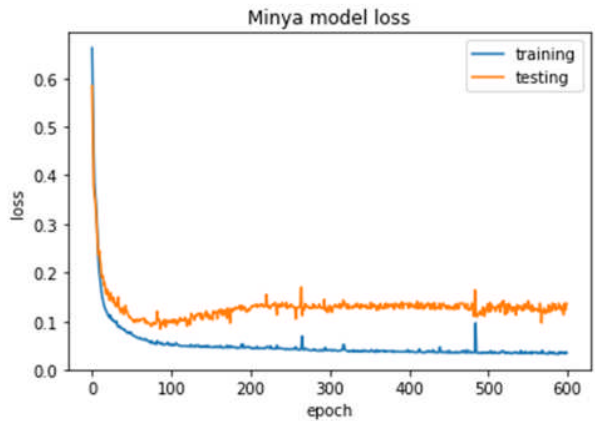

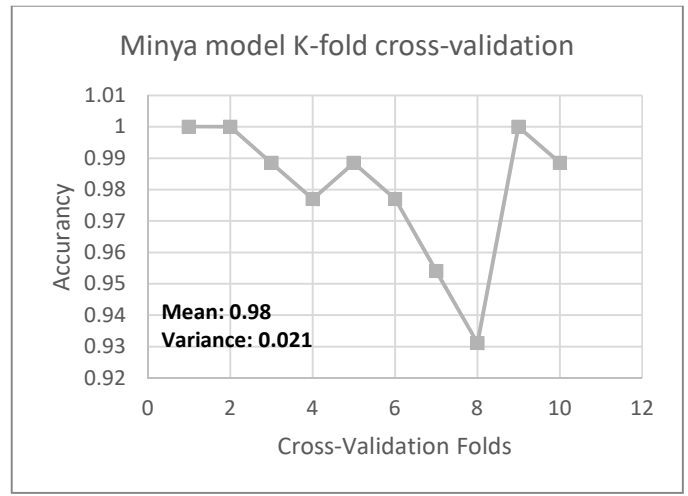

| Confusion matrix | | Actual values | |
|---|---|---|---|
| | | Formal | Informal |
| Predicted | Formal | 400 | 15 |
| | Informal | 14 | 814 |

Total cells: 1243
Overall validation accuracy: 97.6%

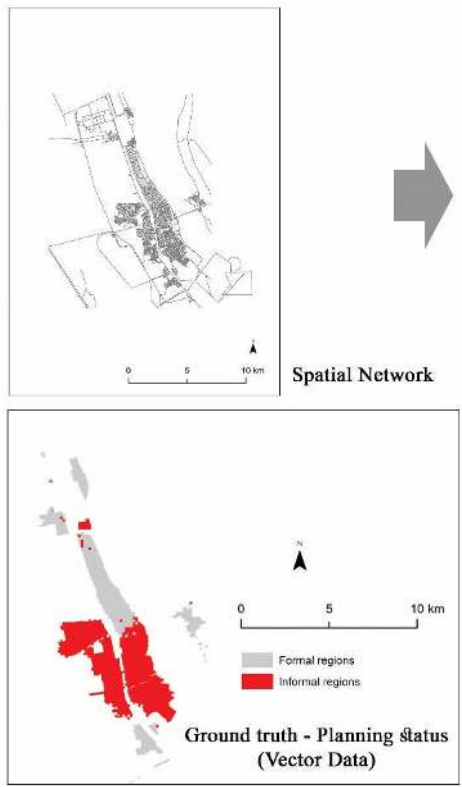

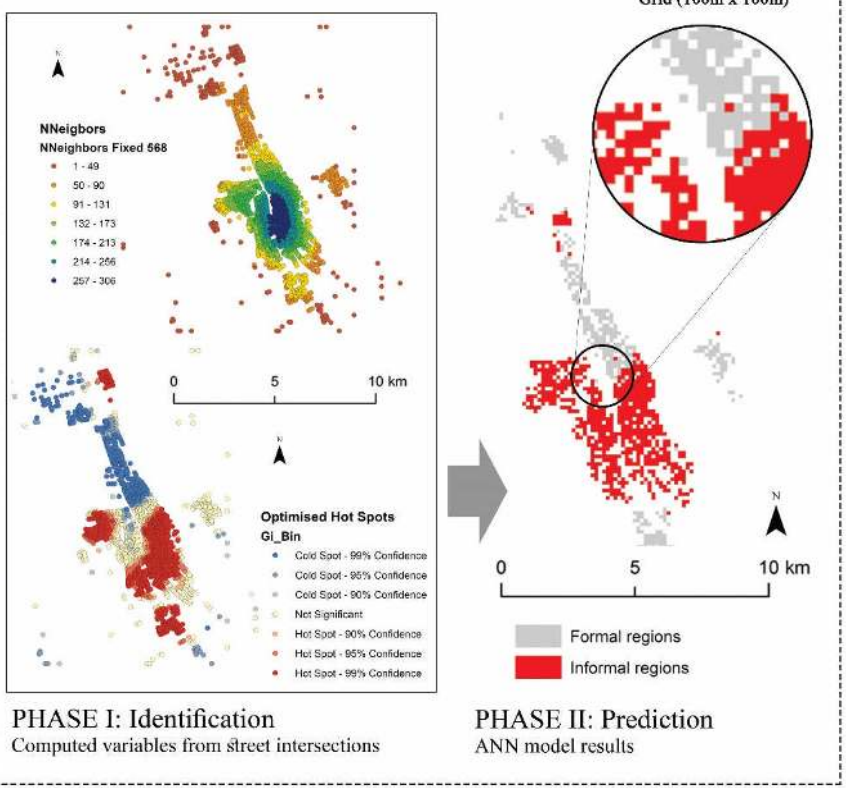

*FIG. 16*
PREDICTSLUMS - MINYA: IDENTIFICATION, TRAINING AND PREDICTION

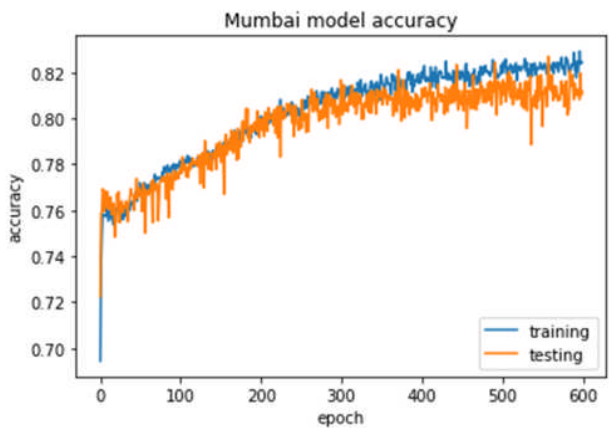
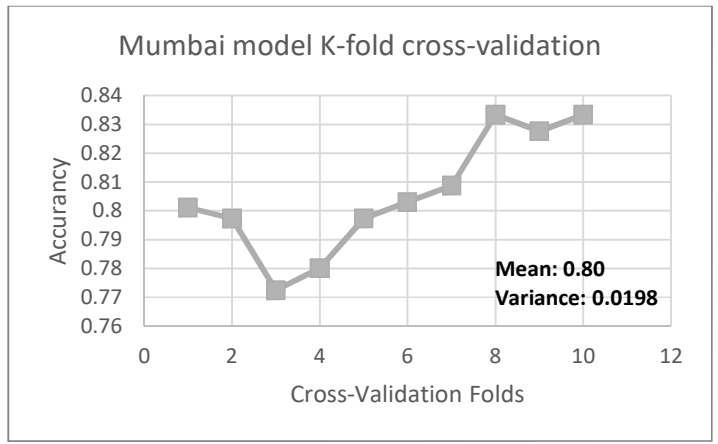
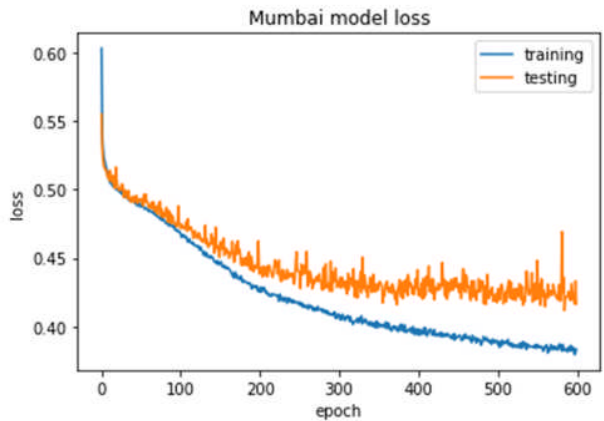

| Confusion matrix | | Actual values | |
|---|---|---|---|
| | | Formal | Informal |
| Predicted | Formal | 2293 | 1059 |
| | Informal | 271 | 3845 |

Total cells: 7468
Overall validation accuracy: 82.1%

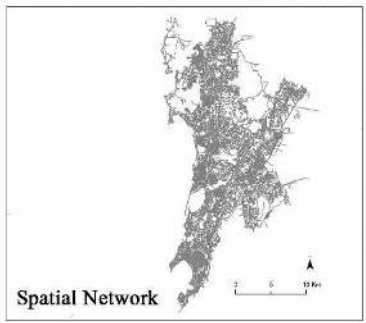
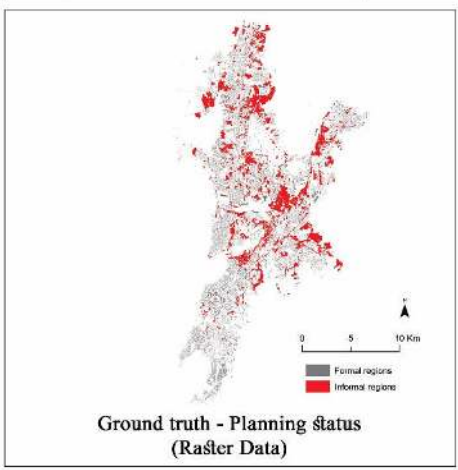
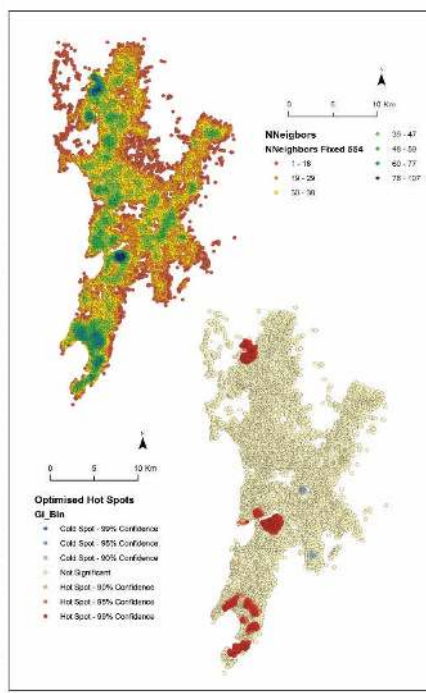
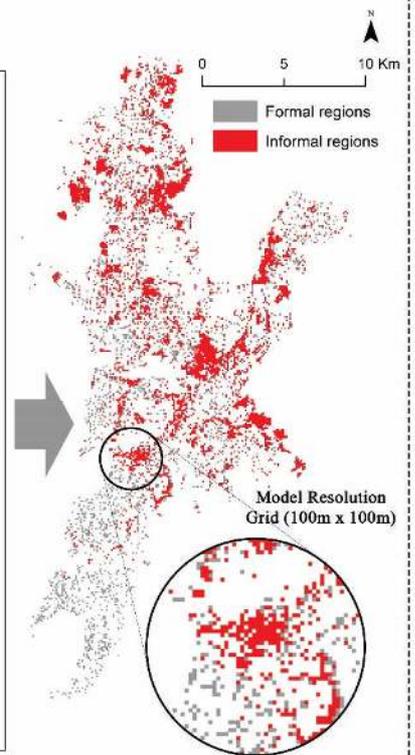

*FIG. 17*
*predictSLUMS - Mumbai: identification, training and prediction*

## 6. DISCUSSION

### 6.1 Informality and scaling laws: number of intersections, street length, and population

Even though the morphologies and the sizes of the studied cities vary, and different urbanization processes may have shaped them, they still follow common principles of growth. This may be one of the reasons that allowed the model algorithms to function in different cities regardless of their spatial profile. In fact, these general scaling laws have manifested in many western cities (Bettencourt, 2013; Cottineau et al., 2017; Isalgue et al., 2007; Masucci, Arcaute, Hatna, et al., 2015). Interestingly, despite the degree of informality that exists in Middle Eastern cities, it seems that they also tend to follow a similar behaviour of growth.

By looking at the distribution of the total number of street intersections in a grid of a side 100m ($\partial$) for the five studied cities, it was found that the intersection points ($n$) follow a scaling law that are ideally represented in an exponential decaying function (See Fig. 18). This finding is in line with that observed in the general distribution of street intersections in London and California (Masucci, Arcaute, Hatna, et al., 2015), although they were represented in a form of an inverse power equation. It is noted that, $n$ for the studied cities is linearly associated with both; the total street length and the population size of the city (See Fig. 19 and Fig. 20).

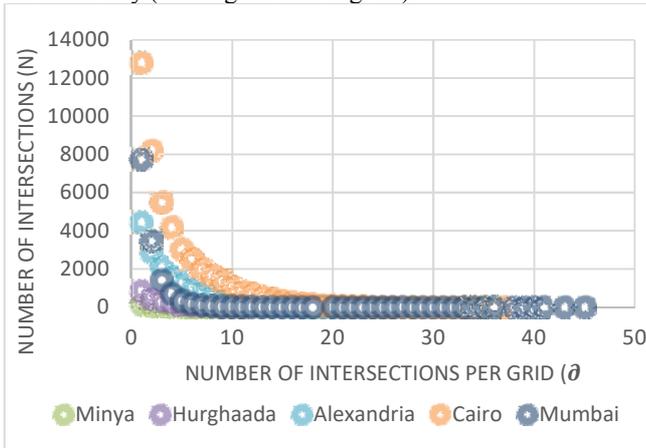

FIG. 18
DISTRIBUTION OF INTERSECTION POINTS

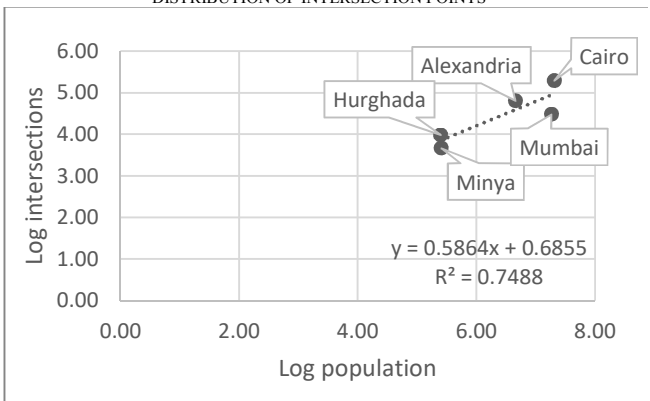

FIG. 19
THE RELATION BETWEEN INTERSECTION POINTS AND POPULATION FOR THE STUDIED FOUR CITIES

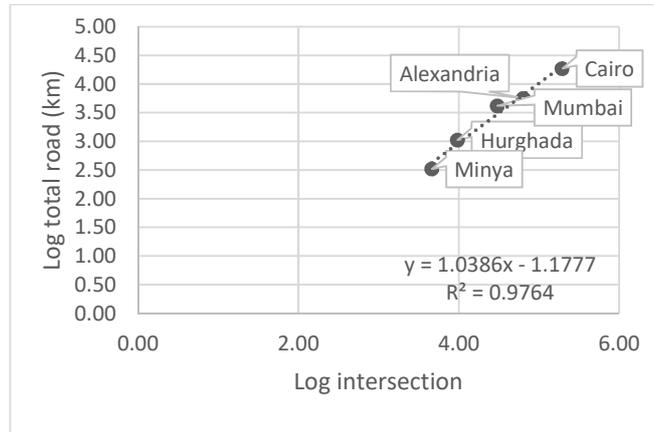

FIG. 20
THE RELATION BETWEEN INTERSECTION POINTS AND LENGTH OF THE TOTAL ROAD FOR THE STUDIED FOUR CITIES

### 6.2 Informality: different forms but similar features?

The main concept and motivation behind the proposed model is to reduce the required input data for classification and prediction of informal regions. This can make it possible for policy-makers and planners to overcome the limitation of poor data availability of informal settlements while still identify informality.

It is crucial to pinpoint that even though informal regions take different forms and shapes based on the local characteristics of the city and its spatial configuration, these informal regions still reserve unique and subtle identifiers that distinguish them from formal settlements. Consequently, this model can detect informal areas and slums in zones that lack data, while retaining accuracy of prediction, by training the model in areas of similar context. However, it should be noted that the accuracy tends to be enhanced when the model is trained in cities that are geographically close to the one for which the prediction is being done.

The representation of hot-cold spots seems to be better in larger cities where informal regions are characterised by urban features, rather than those cities enshrouded with rural characteristics. From the MNL residuals, the model has a better prediction accuracy in Cairo and Alexandria than in Minya and Hurghada, where informal areas are less compact.

### 6.3 Prediction accuracy and spatial resolution

In general, the model shows a good accuracy and validation across the entire datasets of the five studied cities. This opens the door for further application of the model globally. More specifically, the accuracies of each model in each city may vary for several reasons. First, the complexity of cities is relative to their structure and their size which may influence the overall accuracy of predicting informal areas in cities. Second, the consistency of the training and calibrating data used as input is another factor that could influence the overall accuracy of the model. Such factors cannot be controlled entirely in each city. However, by studying more cities of different contexts and with different input data, the learning algorithms of the model can be adapted to deal with these issues. Lastly, the date of the data used for analysis is another crucial aspect that may influence the precision of the output of the model. For instance, the model may accurately predict informality based on the date of the spatial data used, but this can be compromised when it is calibrated and validated using older data on the formal delineation of informal areas.

The spatial resolution of the predictSLUMS model is another key element that influences the accuracy of the model. While prediction of the model is based on the centroids of the grid cells, the current version of the model produces a prediction map in a grid lattice of side 100m. However, the actual prediction of the model depends on input data on planning status, used for training and validation. For example, if the model is trained and calibrated based on a vector based map of informal areas at the land parcel level, the results could be given to the same resolution. However, since not all the data for each city is available in this format, the current version of the model is standardized to a grid lattice of a side 100m, thus giving the opportunity to apply the model in different contexts regardless of the precision of the data used for validation.

### 6.4 Spatial network and OpenStreetMap (OSM) data

While the input data of the model is minimal, there is still variation in the consistency and availability of appropriate spatial data. In this paper, we have relied mainly on the official spatial network data for the Egyptian cities, mainly for consistency and accuracy reasons. However, in order to promote the usage of the model regardless of the type of data, we relied on OSM data for the spatial network for the application of the model in Mumbai. While OSM data has become a ubiquitous source of information, its consistency may vary from city to city. However, there is continuous development of various methods to enhance the acquisition and analysis of complex street networks, i.e. OSMnx (Boeing, 2017b), which suggest that current inconsistencies may soon be a problem of the past.

### 6.5 Limitations

While the predictSLUMS model offers significant improvements for defining the urban system of housing informality in various cities over other techniques, there are still various limitations that need to be addressed in future research.

First, the proposed model is a predictive static model. The challenges remain in introducing a temporal scale to the model. However, before moving into a dynamic model, two steps seem to be essential. First, addressing the third dimension of space and understanding how vertical sprawl takes place, as well as understanding the degrees of informality, seem to be crucial factors for introducing a reliable and useful predictive model in the case of informality.

A second issue remains as to how the model is validated. Currently, it relies on data on well-identified informal settlements. This requires further enhancement to include the different context of missing data. In order to predict future scenes of informality, the model introduces the first steps towards predicating unlabelled data in a city of the currently defined regions by using the pre-trained models. However, the challenges remain in overcoming the paucity of the validation data of the planning status that is required for training and validating the model. A potential way to overcome this issue is to shift the training of the model from an offline way to online training when crowdsourcing data or a potentially developed platform is available for the purpose of updating the planning status of the built environment.

### 6.4 predictSLUMS vs satellite image classification

Currently, methods relying on satellite image classification represent the current state-of-the-art of informal and slums detection. However, there are several drawbacks of these methods (Mahabir et al., 2018), that make the predictSLUMS model a better state-of-the-art for identifying and predicting slums and informal areas. These drawbacks are:

1. Image classification does not define the urban system of informal regions in cities, it is only confined to the specific regions that the study took place in, in which in many cases not even an entire city and cannot be duplicated in other places as the way the model is trained is only valid for a certain case.

2. The data accessibility of such high resolution satellite images or data coming from unmanned vehicles that are commonly used for this approach make such studies only valid for certain regions and cannot be used in others, specifically in the case of Egypt or the MENA region, where is this data will come from? The minimal input data of the predictSLUMS enables it to be used by planners and policy-makers in different regions.

3. from a technical point of view, dealing with points is less computationally expensive than classifying images. This can allow the model to be conducted on a mega scale.

4. In term of model accuracy, the models show good results above 80%, that even more accurate than many studies done in the past using image classification.

5. This model algorithm can deal with different cities, nevertheless, allows the possibility of predicting informal areas in a city from the labelled data of another city.

## 7. CONCLUSION AND FUTURE WORK

Understanding urban systems remains a crucial challenge for planners and policy-makers. It is unequivocal that informal processes enlarge the complexity of understanding the dynamics of cities. This paper aims to contribute to the methods of urban modelling by using machine learning and artificial intelligence in identifying parts of complex systems. It highlights the importance of finding a unifying definition and unique identifiers that can pragmatically represent informal settlements in Egypt and elsewhere, and untangle the complexity of their forms and shapes. The paper focused on answering two research questions. First, how it is possible to infer the status of the built environment from street network data. Second, whether it is possible to identify and predict informality in a city by understanding informality in others.

In this research, we introduced the predictSLUMS model; it is an unprecedented approach to identifying and predicting informal settlements from street intersections. The model algorithms rely on both spatial statistics and machine learning approaches. After computing two variables from street intersections that are likely to represent informality, Multinomial Logistic Regression (MNL) and Neural networks (ANN) have been used to validate and predict informal regions within the same city as the training data, and in a different one. The model's key features can be summarized in three points. First, it requires minimal input data to function. Second, the model can identify and predict hotspots that represents informality in a city. Last, by training the model in one city, it can predict informality in a different city. This minimal requirement of input data enables it to be used by policy-makers and planners in developing countries, where data availability can be a major issue.

The model has been computed for five major cities, including; Greater Cairo, Alexandria, Hurghada, and Minya in Egypt and Mumbai in India. It identifies informal settlements with a high degree of accuracy regardless of their density or degrees of severity. While the overall accuracy of the model is high for identification and prediction, it does vary from city to a city. When using MNL, the model shows a better identification of informal regions in larger cities when compared to smaller ones.

As for future work, three areas seem to be significant for the expansion of this model besides addressing the model limitation:

1) Exploring other contexts globally to enhance the model performance in understanding the subtitles of informal and slum areas in regions beyond the studied five cities.

2) Moving from static to dynamic to not only predict formal/informal areas but also to explore the formation of slums through time.

3) Moving from offline training to online training through an interactive platform where the model validation and calibration can be based on up-to-date evidence.

## 8. ACKNOWLEDGEMENT

The outcome of this research is a part of a PhD study for the first author at University College London, supported by a partial fund of UCL Overseas Research Scholarship (ORS). We would like to thank Prof. Peter Jones for his feedback and suggestions towards improving this article.